\documentclass[reprint, superscriptaddress, aps, prb, nofootinbib, balancelastpage]{revtex4-2}
\usepackage{graphicx} % Required for inserting images
\usepackage{tikz}
\usepackage{hyperref}
\usepackage{amsmath}
\usepackage{mathtools}
\usepackage{amssymb}
\usepackage{bm}
\usepackage{mdframed}
\usepackage[normalem]{ulem}
\usepackage{acro}
\usepackage{chemformula}

\DeclareAcronym{DFT}{
short = DFT,
long = Density-Functional Theory
}

\DeclareAcronym{LE}{
short = LE,
long = Laplcaian Eigenfunctions
}

\DeclareAcronym{SOAP}{
short = SOAP,
long = Smooth Overlap of Atomic Positions
}

\DeclareAcronym{ACE}{
short = ACE,
long = Atomic Cluster Expansion
}

\DeclareAcronym{MACE}{
short = MACE,
long = Multi-ACE
}

\DeclareAcronym{RR}{
short = RR,
long =  Ridge Regression
}

\DeclareAcronym{KRR}{
short = KRR,
long = Kernel Ridge Regression
}

\DeclareAcronym{GAP}{
short = GAP,
long = Gaussian Approximation Potential
}

\DeclareAcronym{II}{
short = II,
long = Information Imbalance
}

\DeclareAcronym{RP}{
short = RP,
long = Random Projection
}

\DeclareAcronym{JL}{
short = JL,
long = Johnson-Lindenstrauss
}

\DeclareAcronym{TR}{
short = TR,
long = tensor-reduced
}

\DeclareAcronym{MPNN}{
short = MPNN,
long = Message Passing Neural Network
}

\DeclareAcronym{CLSR}{
short = CLSR,
long = Compressed Least-Square Regression
}

\DeclareAcronym{MD}{
short = MD,
long = Molecular Dynamics
}

\DeclareAcronym{MLIP}{
short = MLIP,
long = machine-learned interatomic potential
}

\DeclareAcronym{PES}{
short = PES,
long = Potential Energy Surface
}

\DeclareAcronym{NEB}{
short = NEB,
long = Nudged Elastic Band
}

\DeclareAcronym{RSS}{
short = RSS,
long = Random Structure Searching
}

\hypersetup{
    colorlinks=true,
    linkcolor=blue,
    filecolor=magenta,      
    urlcolor=blue,
    citecolor=blue
    }

\def\rr{{\bm r}}

\def\bAA{{\bm A}}

\def\ll{{\bm l}}
\def\nn{{\bm n}}
\def\mm{{\bm m}}

\def\cc{{\bm c}}

\def\rcut{r_{\rm cut}}

\def\yy{{\bm y}}

\newcommand*\diff{\mathop{}\!\mathrm{d}}

\begin{document}

\title{Regularity Priors for the Linear Atomic Cluster Expansion }

\author{James P. Darby}
\affiliation{Engineering Laboratory, University of Cambridge, Cambridge, CB2 1PZ UK}
\author{Joe D. Morrow}
\affiliation{Inorganic Chemistry Laboratory, Department of Chemistry, University of Oxford, Oxford OX1 3QR, UK}
\author{Albert P. Bartók}
\affiliation{Department of Physics and Warwick Centre for Predictive Modelling, School of Engineering,  University of Warwick, Coventry, CV4 7AL, UK}
\author{Volker L. Deringer}
\affiliation{Inorganic Chemistry Laboratory, Department of Chemistry, University of Oxford, Oxford OX1 3QR, UK}
\author{G\'abor Cs\'anyi}
\affiliation{Engineering Laboratory, University of Cambridge, Cambridge, CB2 1PZ UK}
\author{Christoph Ortner}
\affiliation{Department of Mathematics, University of British Columbia, 1984 Mathematics Road, Vancouver, BC, Canada V6T 1Z2}

\date{\today}

\begin{abstract}
% CO: SHORTENED ABSTRACT
%
Machine-learned interatomic potentials enable large systems to be simulated for long time scales at near \textit{ab-initio} accuracy. This accuracy is achieved by fitting extremely flexible model architectures to high quality reference data. In practice, this flexibility can cause unwanted behavior such as jagged predicted potential energy surfaces and generally poor out-of-distribution behavior. We investigate a general strategy for incorporating prior beliefs on the regularity of the target energy into linear Atomic Cluster Expansion (ACE) models and explore to what extent this approach improves the quality of the fitted models. Our main focus is an over-regularisation that replicates the Gaussian broadening used in Smooth Overlap of Atomic Positions (SOAP) descriptors within the ACE framework.
% We show that the Gaussian broadening used in SOAP descriptors can be replicated with a specific choice of ``regularity prior''.
% Various forms for the ``regularity prior'' are investigated and we show that the Gaussian broadening used in SOAP descriptors can be replicated with a specific choice of prior. 
Numerical tests indicate that the exact form of the prior is non-critical but that including such a prior leads to significant improvement in test errors, consistent repulsion at close-approach, eliminates spurious false minima in the potential energy and enhances stability during molecular dynamics simulations. 
% Due to their diagonal forms the priors can be used with standard solution methods and incur no additional computational cost during fitting.
%
\end{abstract}

% BACKUP ABSTRACT 
% The development of machine-learned interatomic potentials has transformed computational materials science, enabling large systems to be simulated for increasingly long time scales at near \textit{ab-initio} accuracy. This accuracy is achieved by fitting extremely flexible architectures with many learned parameters to high quality reference data. In practice, this flexibility can cause unwanted behavior such as jagged predicted potential energy surfaces, with spuriously sharp features and poor extrapolation outside of the training domain. In this work, we investigate how prior beliefs on the smoothness of the energy can be incorporated into linear ACE models via a modified regularization term. Various forms for the ``regularity prior'' are investigated and we show that the Gaussian broadening used in SOAP descriptors can be replicated with a specific choice of prior. Numerical tests indicate that the exact form of the prior is non-critical but that including such a prior leads to significant improvement in test errors, helps achieve consistent repulsion at close-approach, helps eliminate spurious false minima in the potential energy surface and enhances stability during molecular dynamics simulations. Due to their diagonal forms the priors can be used with standard solution methods and incur no additional computational cost during fitting.

\maketitle

\section{Introduction}

For the past few decades \ac{DFT} \cite{hohenberg1964inhomogeneous, kohn1965self} has been the workhorse of computational chemistry, however, the cubic scaling with the number of electrons places significant restrictions on the accessible simulation sizes and timescales. 
In contrast, classical force fields \cite{muser2023interatomic} offer linear scaling and are fast to evaluate, but, the simple functional forms used limit their accuracy and typically lead to a poor description of complex processes such as fracture \cite{zhang2023atomistic}, phase transitions \cite{erhard2022machine, marchant2023exploring}, or chemical reactions \cite{mattsson2010first, heijmans2020gibbs}.
\Acp{MLIP} also approximate the total energy as a sum of atomic energies \cite{behler2007generalized}, but do so using highly flexible architectures often involving millions of learned parameters.
As such, provided they are trained on a sufficient quantity of high fidelity data, they can provide near \textit{ab-initio} accuracy across a much wider range of simulation scenarios.

% For the past few decades \ac{DFT} has been the workhorse of computational chemistry, however, the cubic scaling with the number of electrons places significant restrictions on the accessible simulation sizes and timescales. 
% Classical force fields coarse grain away the electrons and approximate the total energy as a sum of atomic energies, each of which is predicted based on the local environment around the atom of interest. 
% This leads to  linear scaling with system size and makes these models blazingly fast to evaluate. 
% However, the downside of traditional empirical potentials is that they use relatively simple functional forms, with a handful of parameters that are typically fitted to reproduced material properties  across a relatively narrow range of conditions \cite{muser2023interatomic}. 
% This limits their accuracy and leads to limited transferability which often manifests itself as a poor description of events such as fracture \cite{zhang2023atomistic}, phase transitions \cite{erhard2022machine, marchant2023exploring}, or chemical reactions \cite{mattsson2010first, heijmans2020gibbs}.  
% In contrast, \acp{MLIP} model the site energies using highly flexible architectures, often with millions of learned parameters. 
% As such,  provided they are trained on a sufficient quantity of high fidelity data, they can provide near  \textit{ab-initio }accuracy across a much wider range of configurations whilst retaining the linear scaling with system size.

However, the extreme flexibility of ML architectures has drawbacks. In particular, MLIPs are notoriously poor at extrapolating outside the domain of the training dataset \cite{vita2023data, benoit2020measuring}, which often leads to catastrophic failure during \ac{MD} simulations e.g. a ``hole'' in the \ac{PES} could be encountered causing the predicted energy to decrease by $\sim$100s-1000s of eV/atom and the simulated system to explode \cite{nandi2019using, van2020regularised}. Avoiding such behavior tends to require careful construction of the training dataset, either using labour intensive manually chosen configurations, or via automated approaches such as active learning \cite{van2023hyperactive, sivaraman2020machine}. The recent wave of foundation potentials~\cite{batatia2023foundation, yang2024mattersim, neumann2024orb, park2024scalable, fu2025learning, barroso2024open, merchant2023scaling, rhodes2025orb, bochkarev2024graph} have come a long way towards fixing this issue, primarily by using enormous training datasets in combination with careful architectural decisions. These potentials show unprecedented stability across a diverse range of materials and simulations \cite{batatia2023foundation} whilst retaining meaningful accuracy. However, they are not perfect, with the occasional hole still encountered for particularly unusual configurations or, more commonly, for high pressure configurations under extreme compression where there is relatively little training data available. 
Furthermore, some architectures are capable of achieving excellent point wise errors on energies, forces and stresses \cite{riebesell2025framework} whilst interpolating between the training data in a surprisingly ``rough'' fashion, with the predicted \ac{PES} showing small amplitude high frequency oscillations, which can be sufficiently sharp to appear almost as kinks. This behavior is seen in the homo-nuclear diatomic energy curves presented in the MLIP Arena benchmark \cite{chiang2025mlip} and can lead to many practical concerns, such as failure of geometry optimization or transition path search~\cite{henkelman2000climbing} or encountering false minima during crystal structure prediction \cite{deringer2018data}.

\begin{figure}[h!]
    \centering
    \begin{tikzpicture}
        \node[anchor=south west, inner sep=0] (img) 
            {\includegraphics[width=0.5\textwidth]{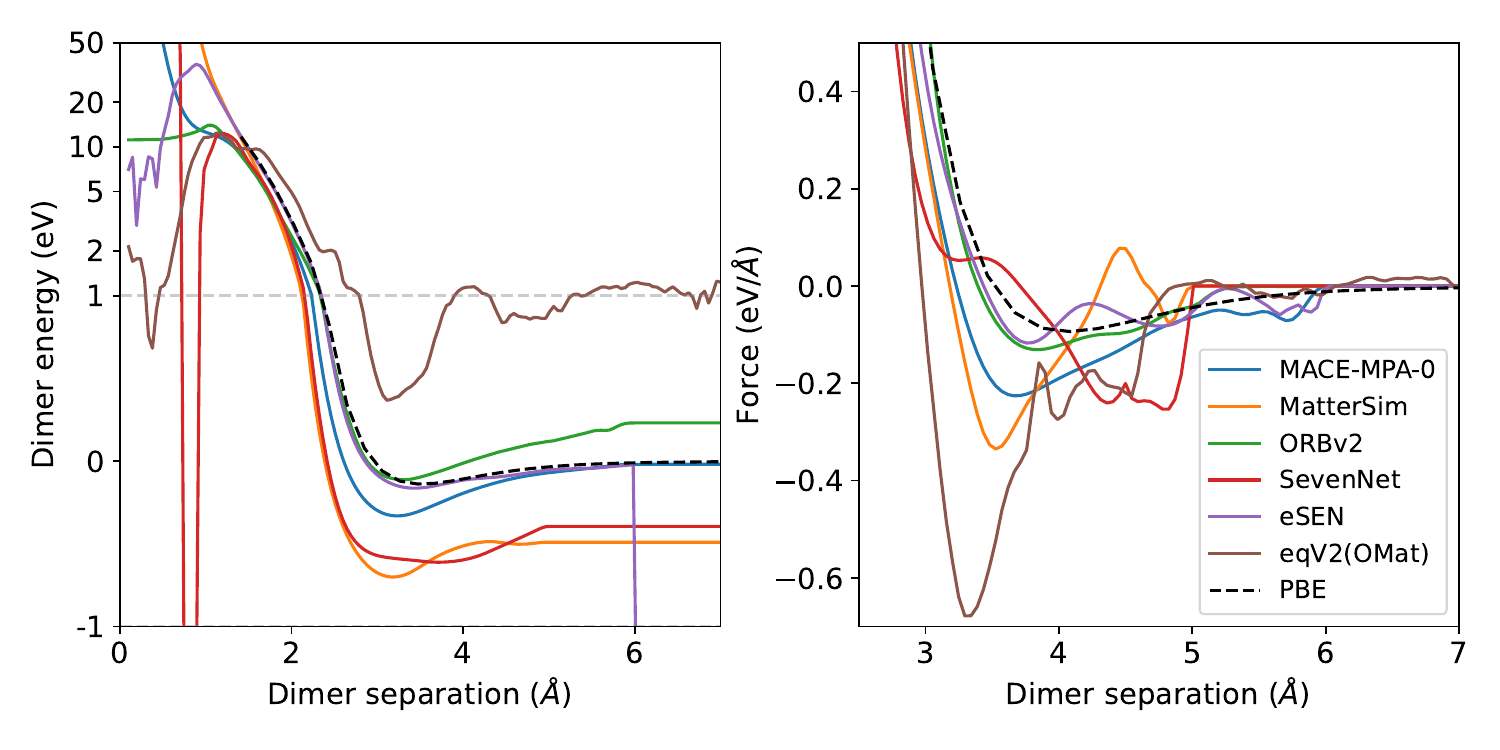}};
        
        \begin{scope}[x={(img.south east)}, y={(img.north west)}]
            \node[font=\small] at (0.45,0.87) {(a)};
             \node[font=\small] at (0.93, 0.87) {(b)};
        \end{scope}
    \end{tikzpicture}
    \caption{The energy (a) and force (b) predictions on the Mg-Mg dimer are shown for various foundation models \cite{batatia2023foundation, yang2024mattersim, neumann2024orb, park2024scalable, fu2025learning, barroso2024open} as a function of dimer separation. The MLIP Arena python package \cite{chiang2025mlip} was used to load all models and obtain the reference PBE result, originally computed with VASP \cite{kresse1996efficiency}. For the avoidance of doubt, model versions are MACE-MPA-0, MatterSim-v1.0.0-5M, ORB-v2, 7net-0, eSEN-30M-OAM and eqV2\_86M\_omat\_mp\_salex.}
    \label{fig:Mg_dimers}
\end{figure}

% \begin{figure}[h!]
%     \centering
%     \includegraphics[width=0.5\textwidth]{Figures/MLIP_arena_Mg_dimer.pdf}
%     \caption{The energy and force predictions on the Mg-Mg dimer are shown for various foundation models \cite{batatia2023foundation, yang2024mattersim, neumann2024orb, park2024scalable, fu2025learning, barroso2024open} as a function of dimer separation. The MLIP Arena python package \cite{chiang2025mlip} was used to load all models and obtain the reference PBE result, originally computed with VASP \cite{kresse1996efficiency}. For the avoidance of doubt, model versions are MACE-MPA-0, MatterSim-v1.0.0-5M, ORB-v2, 7net-0, eSEN-30M-OAM and eqV2\_86M\_omat\_mp\_salex. }
%     \label{fig:Mg_dimers}
% \end{figure}

The issues described above are ``obvious'' failings in the sense that we can identify them almost at a glance for two key reasons. Firstly, empirically, the reference \ac{PES} in most low-energy regions of configuration space for chemical and materials systems is smooth in the sense that it varies over a characteristic length scale that depends on the size of the atoms in questions, for example see figure\ref{fig:Mg_dimers}.
Secondly,the energy should increase sharply at short interatomic distances $r$ due to a combination of Pauli and Coulomb repulsion, with the $1/r$ Coulomb repulsion between nuclei dominating at very close approach \cite{ziegler1985stopping, nordlund2025repulsive}.
In this work we investigate how this chemical intuition can be incorporated into linear \ac{ACE} models by adjusting the form of the regularization term. We focus on ACE for two reasons: (i) it remains an attractive light-weight alternative \cite{zhou2025full} to large-scale message passing neural networks; and (ii) it provides a prototype study that can be extended to many related architectures, such as MACE \cite{batatia2022mace} and GRACE \cite{bochkarev2024graph}, that build on the ACE framework. 

When fitting linear models the standard choice is to employ Tikhonov regularization. This leads to a single parameter, the regularization strength, which balances model accuracy (on the training set) against stability, by penalizing models with large coefficients. From a Bayesian perspective, such a choice implies a uniform prior belief for the coefficients of all basis functions, irrespective of their polynomial degree. Here, we investigate different ``shapes'' for the regularization which aim to incorporate our prior belief on the smoothness of the PES. By smoothness, or more formally regularity, we not only mean that the energy should be continuous and differentiable, but also that it should not contain small amplitude high-frequency oscillations. This is analogous to the length-scale parameter within Gaussian Process Regression kernels \cite{bartok2010gaussian, deringer2021gaussian} and we show how the $\sigma$ parameter in \ac{SOAP} descriptors \cite{bartok2013representing} corresponds to a particular choice of regularization within linear ACE.  In practice, we find that the exact form of the regularization term is not critical, but that picking one which incorporates a notion of regularity leads to significant improvement in model performance without incurring any additional computational cost. We validate this with a variety of numerical tests including error estimates, \ac{MD} stability, and \ac{RSS} as well as by comparing various cuts through the predicted \ac{PES} with the true reference. 

\section{Methods}

\subsection{Regularized Atomic Cluster Expansion}
Within the basic atomic cluster expansion framework~\cite{drautz2019atomic,dusson2022atomic} an atom is represented as $(\rr_i, Z_i)$ where $\rr_i$ is the position and $Z_i$ a categorical variable specifying the chemical element. Relative positions are denoted by $\rr_{ij} = \rr_i - \rr_j$ and ``bonds'' between atoms $i, j$ by $x_{ij} = (\rr_{ij}, Z_i, Z_j)$.  The ACE site energy is defined in terms of a self-interacting many-body expansion. Let $\mathcal{N}(i)$ be the atomic environment of atom $i$, i.e. all atoms within distance $r_{ij} \leq \rcut$, then 
\begin{equation}
    \begin{aligned}
    E_i &= E_i\big( \{x_{ij}\}_{j \in \mathcal{N}(i)} \big) \\ 
    &= \sum_{N = 0}^{N_{\rm max}}
      \sum_{j_1, \dots, j_N \in \mathcal{N}(i)}  
      U_N(x_{i j_1}, \dots, x_{i, j_N}).
  \end{aligned}
\end{equation}

Each $(N+1)$-body potential $U_N$ is expanded in terms of tensor products of the one-particle basis, resulting in 
\begin{equation}
    U_N(x_{i j_1}, \dots, x_{i j_N} ) 
    =  
    \sum_{ \nn \ll \mm} 
    c_{\nn \ll \mm}  \prod_{t=1}^{N}\phi_{n_t l_t m_t}(x_{i j_t}).     
\end{equation}
where the one-particle basis $\phi_{nlm}$ acts on bonds $x_{ij}$ and is defined by 
\[
    \phi_{nlm}(\rr_{ij}, Z_i, Z_j) 
    = 
     R_{nl}(r_{ij}, Z_i, Z_j) Y_l^m(\hat{\rr}_{ij}),
\]
where $R_{nl}$ is called the {\em radial basis} and $Y_l^m$ are the (real) spherical harmonics. For notational convenience there is not a specific index for the chemical element, instead the element dependence is absorbed into the $n$ index on $R_{nl}$, see Eq. \ref{eq:ACEpot_radials}. 
The tensor product structure of the summation over $j_t$-tuples can be exploited to rewrite $E_i$ in a form that is highly efficient computationally \cite{bartok2013representing, drautz2019atomic},
\begin{equation}
    \label{eq:ACE1}
    \begin{aligned} 
        A_{i, n l m} 
        &= \sum_{j \in \mathcal{N}(i)} 
        \phi_{ n l m}(x_{i j}), 
        \\ 
        \bAA_{i, \nn \ll \mm}
        &= \prod_{t = 1}^N A_{i,  n_t l_t m_t} \\ 
        E_i 
        &= \cc \cdot \bAA = 
        \sum_{ \nn \ll \mm} 
        c_{ \nn \ll \mm} \bAA_{i, \nn \ll \mm},
    \end{aligned}
\end{equation}
so that the evaluation cost is now approximately linear in the number of neighboring atoms and linear in the number of parameters. 

% ----------------------------------------
% B = C * A 
% cbar' * B = c' * A = cbar' * C * A 
%           = (C' * cbar)' * A
% => c = C' * cbar
%
% Constraint: c in range(C')
% If C' has orthogonal columns, then 
%      c = C' * C * c
%      (I - C' * C) * c = 0
% ----------------------------------------

Rotational, or $O(3)$, invariance is imposed by forcing the coefficients to obey a linear constraint,
\begin{equation} \label{eq:O3invariance}
    \mathcal{L} {\bm c} = 0.
\end{equation}
This constraint is most commonly implemented through a transformation $\cc = \mathcal{C}^T \overline{\cc}$, which is equivalent to a basis transformation ${\bm B}_i = \mathcal{C} \bAA_i$ with $\mathcal{L} = I - \mathcal{C}^T \mathcal{C}$; see Appendix, Section ~\ref{sec:symbasis_priors} for details. For our purposes it will be clearer to work with the $\bAA$ basis and simply keep the constraint \eqref{eq:O3invariance} in mind. 

Being a linear model, the problem of finding the coefficients using regularized least squares regression is of the form, 
\begin{equation}
   \min_\cc  \| X \cc - \yy \|^2 + \lambda \| \Gamma \cc \|^2, 
\end{equation}
subject to \eqref{eq:O3invariance}, where $X, \yy$ is given by the fitting data, $\lambda$ is a tuning parameter and the Tikhonov operator $\Gamma$ will be chosen diagonal and of the form 
\begin{equation}
    \Gamma_{ \nn \ll \mm,  \nn \ll \mm}
    = \gamma_{ \nn \ll}.
\end{equation}
We are foremost concerned with different choices for $\Gamma$, how they can be interpreted and how they affect model performance and predicted properties. Specifically, we will explain how certain choices of $\gamma$ can be interpreted as regularity priors for the potentials $U_N$ and, by investigating the effect of each prior on the implied atomic neighbor density, draw a connection to the Gaussian broadening used when combining \acp{GAP} with SOAP descriptors\cite{klawohn2023gaussian}.

Note that we do not allow $\mm$-dependent $\gamma$ as this would imply a preferred orientation; see section \ref{sec:neighbour_density} for more details. 

\subsection{Regularity Priors}
The Tikhonov operator $\Gamma$ can also be interpreted within the Bayesian framework as a prior on the parameters; $\cc \sim \mathcal{N}({\bm 0}, \Gamma^{-2})$. Since the parameters $\cc$ are directly related to the expansion of $U_N$, the operator $\Gamma$ therefore encodes our ``prior belief'' about those potentials. A natural class of priors are regularity priors. The idea is that, if $\phi_{ nlm}$ is an orthogonal polynomial basis (in a suitably chosen coordinate system), then the $(N+1)$-body potentials $U_N$ can be expanded {\em exactly} as 
\begin{equation}
    U_N = \sum_{\nn\ll\mm} c_{\nn\ll\mm} 
    \bigotimes_{t=1}^N \phi_{ n_t l_t m_t}.
\end{equation}
The characteristic decay of the coefficients $c_{\nn\ll\mm}$ is directly related to the regularity class to which $U_N$ belongs. Roughly speaking, the more regular (``smooth'') a function is the faster the coefficients decay.  Sharp characterizations of this decay in dimension greater than one are technically subtle and depend on fine details of the interplay between the choice of basis $R_{nl}$ and precise characterizations of multi-variate regularity of $U_N$ which goes beyond the scope of this work. Roughly speaking, one can expect the following behavior on two prototypical cases: 
\begin{equation} \label{eq:decay_Cp_analytic}
    c_{\nn\ll\mm} \lesssim 
    \begin{cases} 
        \prod_t \big(1 + \sigma_{\rm{n}} n_t + \sigma_{\rm{l}}l_t)^{-p}, & 
         % \hspace{-0.3cm} 
         \text{if $p$ times diff.,} \\ 
        \exp\Big( - \sum_t  \sigma_{\rm{n}} n_t  + \sigma_{\rm{l}}l_t \Big), & 
        \text{if analytic},
    \end{cases}
\end{equation}
where $\sigma_{\rm{n}}$ and $\sigma_{\rm{l}}$ are  parameters.  By $U_N$ being $p$ times differentiable, we mean that $U_N$ has partial derivatives of the form $\partial_{\rr_1}^{p_1} \cdots \partial_{\rr_N}^{p_N} U_N$ up to $\max_t p_t \leq p$.
By $U_N$ being analytic we mean jointly analytic in all variables.
We are again ignoring the $m$ indices since we always have $|m_t| \leq l_t$. 

If our prior belief (or knowledge) is that the $U_N$ belong to one of those classes, then we can encode this into the prior distribution on the parameters by choosing 
\begin{equation}
    \gamma(\nn,\ll) 
    = 
    \begin{cases} 
        \prod_t \big(1 + \sigma_{\rm{n}} n_t  + \sigma_{\rm{l}}l_t \big)^p, & 
        % \hspace{-0.3cm} 
        \text{if $p$ times diff.,} \\ 
        \exp\Big( \sum_t \sigma_{\rm{n}} n_t  + \sigma_{\rm{l}}l_t \Big), & 
        \text{if analytic}. 
    \end{cases}    
\end{equation}

In Bayesian modelling, one generally prefers a ``cautious'' prior that under-regularizes, rather than an over-confident prior that over-regularizes. In our context, this would result in choosing a lower regularity class. 
This rule-of-thumb is based on the assumption that there is sufficient data to specify the model. In the setting of MLIPs this is almost never the case. Even the most diverse datasets only cover a relatively small part of configuration space that is ``likely to be visited often'' during inference (e.g., \ac{MD} simulation). To ensure maximal robustness of the fitted models when extensive sampling of configuration space is required, it is therefore interesting to explore the effect of over-regularization, i.e., selecting regularity priors that may be theoretically too strong. In that spirit we will also consider the Gaussian prior, 
\begin{equation}  \label{eq:gauss_prior}
    \gamma(\nn, \ll)      =      \exp\Big( \sum_t \sigma^2_{\rm{n}} n_t^2    +  \sigma^2_{\rm{l}}l_t^2      \Big). 
\end{equation}
The regularity class of functions having the analogous decay of coefficients is extremely small (the algebra of polynomials and Gaussians), hence this choice should be thought of as over-regularization and not as describing the regularity of the underlying target function. A detailed discussion on the interplay between regularity priors and symmetrization over the rotation group is given in section \ref{sec:symbasis_priors}.

\subsection{Regularization vs Rescaling}
\label{sec:rscaling}
It is always possible to transform $\tilde{\cc} = \Gamma \cc$ so that the effective prior for $\tilde{\cc}$ will be the standard Gaussian $\mathcal{N}({\bm 0}, I)$. This has the advantage that standard parameter estimation methods that implicitly assume the ``canonical'' Tikhonov regularization can be used. For many choices of $\gamma$ the prior can also be interpreted as a rescaling of the basis functions. The priors considered above all have the form

\[
    \gamma( \nn \ll \mm) 
    = 
    \prod_t \gamma_{n_t l_t}. 
\]
For such priors, the ACE model for the site energy transforms as 
\begin{align*}
    E_i 
    &= 
    \cc \cdot \bAA = \tilde\cc \cdot \big( \Gamma^{-1} \bAA \big) \\ 
    &= 
    \sum_{ \nn \ll \mm} 
    \tilde{\cc}_{ \nn \ll \mm} 
    \prod_t \gamma_{n_t, l_t}^{-1} A_{i,  n_t l_t m_t} 
    \\     
    &=: 
    \sum_{ \nn \ll \mm} 
    \tilde{\cc}_{ \nn \ll \mm} 
    \prod_t \tilde{A}_{i,  n_t l_t m_t},
\end{align*}
where $\tilde{A}_{i,  n_t l_t m_t}$ is the rescaled atomic basis, 
\begin{equation}
    \tilde{A}_{i,  n_t l_t m_t} = \gamma_{n_t l_t}^{-1} A_{i  n_t l_t m_t}.
\end{equation}

If the radial basis has an $l$-channel then one can take this further and implement the regularity prior as a scaling of the radial basis alone via 

\begin{equation*}
    \tilde{R}_{nl}(r_{ij}, Z_i, Z_j) 
    = \gamma_{nl}^{-1} R_{nl}(r_{ij}, Z_i, Z_j). 
\end{equation*}
In other words, the regularity prior can equivalently be interpreted as a specific scaling of either the one-particle basis or the radial basis $R_{nl}$. These observations are of practical interest for implementing ACE models, and will also be important in the next step where we draw a connection with Gaussian broadening of the atomic density in the SOAP descriptor. Furthermore, we note that rescaling the basis functions provides a natural way to include such priors in non-linear models where there is no explicit regularised least squares problem.

\subsection{Implied Neighbor Density}
\label{sec:neighbour_density}
ACE features can be understood in terms of symmetrized tensor products of a fictitious atomic neighbor density formed by representing each neighboring atom as a delta function. The neighbour density can be written as 

\begin{align}
    \rho_{i}^{(0)}(\rr) &= \sum_{j \in \mathcal{N}(i)} \delta(\rr-\rr_j) \\
    & = \sum_{ nlm} A_{i, nlm} \phi_{ nlm}(\rr)
\end{align}
where here the $A_{i, nlm}$ are expansion coefficients. As such, scaling the expansion coefficients according to a specific regularity prior $\tilde{A}_{i,  n_t l_t m_t} = \gamma_{n_t l_t}^{-1} A_{i  n_t l_t m_t} $ can be interpreted as modifying the corresponding neighbour density, $\tilde{\rho}$. This connection is particularly interesting since GAP potentials, which use $N=2$ ACE features where each atom is represented by a Gaussian of width $\sigma$, have empirically shown better stability than ACE potentials fit to the same dataset \cite{morrow2022indirect}. To examine this relationship further, let us assume that $\phi_{ nlm}$ are Laplacian eigenfunctions, i.e., are $L^2$-orthonormal and satisfy $- \Delta \phi_{ nlm} = \lambda_{nl} \phi_{ nlm}$ as well as $\phi_{ nlm}(|\rr|=r_\mathrm{cut}) = 0$. Note that $\lambda_{nl}$ has no $m$ dependence as $\Delta$ is isotropic. Upon writing $\gamma_{nl}^{-1} = f(\lambda_{nl})$, we then expand a general modified density

\begin{align*}
    \tilde{\rho}_i &= \sum_{ nlm} \tilde{A}_{i, nlm} \phi_{ nlm} \\ 
    &= 
    \sum_{ nlm} A_{i, nlm} f(\lambda_{nlm}) \phi_{ nlm}  \\ 
    &= 
    \sum_{ nlm} A_{i, nlm}
     f(-\Delta) \phi_{ nlm} \\ 
    &= 
    f(- \Delta) \rho_i^{(0)}.
\end{align*}
In principle, any operator could be used in place of $\Delta$ but, as noted by Bigi et. al. \cite{bigi2022smooth}, the Laplacian is a natural choice as. Specificially, its eigenstates are constructed to minimize their average squared gradient, subject to orthonormality constraints, with $\lambda_{nl}$ quantifying this measure of smoothness. Importantly for our context, powers of $\Delta$ provide the means of specifying regularity classes. For example, one could minimize curvature by considering eigenfunctions of the Bi-Laplacian $\Delta^2$ or one could measure smoothness with respect to a scaled radial coordinate. Here, we choose to use $\Delta$ as i) $\lambda_{nl}$ scales approximately as $\sim  n^2 + l^2$, which allows for a simple correspondence with our regularity priors, and ii) it enables approximate expressions for $\tilde{\rho}$ to be derived. This allows us to better understand the effects of the different priors and make a connection to the Gaussian neighbour density used in SOAP.

For the algebraic regularity prior
\[
    \gamma_{nl}^{-1}   \approx (1 + n + l)^{-p}   \approx \lambda_{nl}^{-p/2}
\]
which results in 
\begin{equation}
    \tilde{\rho}  
    =
    (-\Delta)^{-p/2} \rho_i^{(0)}.
\end{equation}

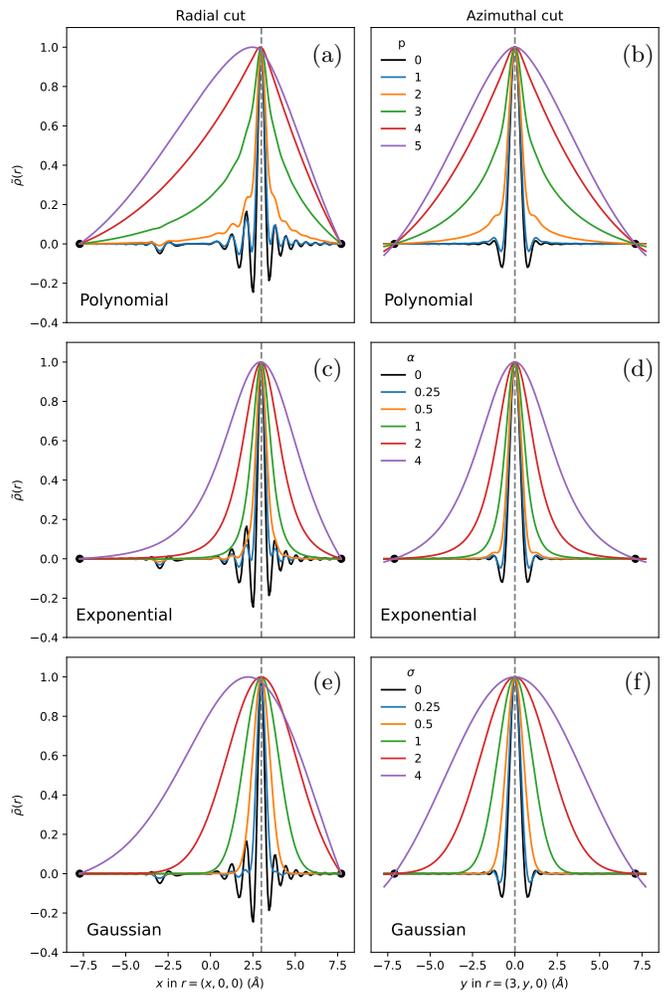
\begin{figure}[h!]
    \centering
    \begin{tikzpicture}
        \node[anchor=south west, inner sep=0] (img) 
            {\includegraphics[width=0.5\textwidth]{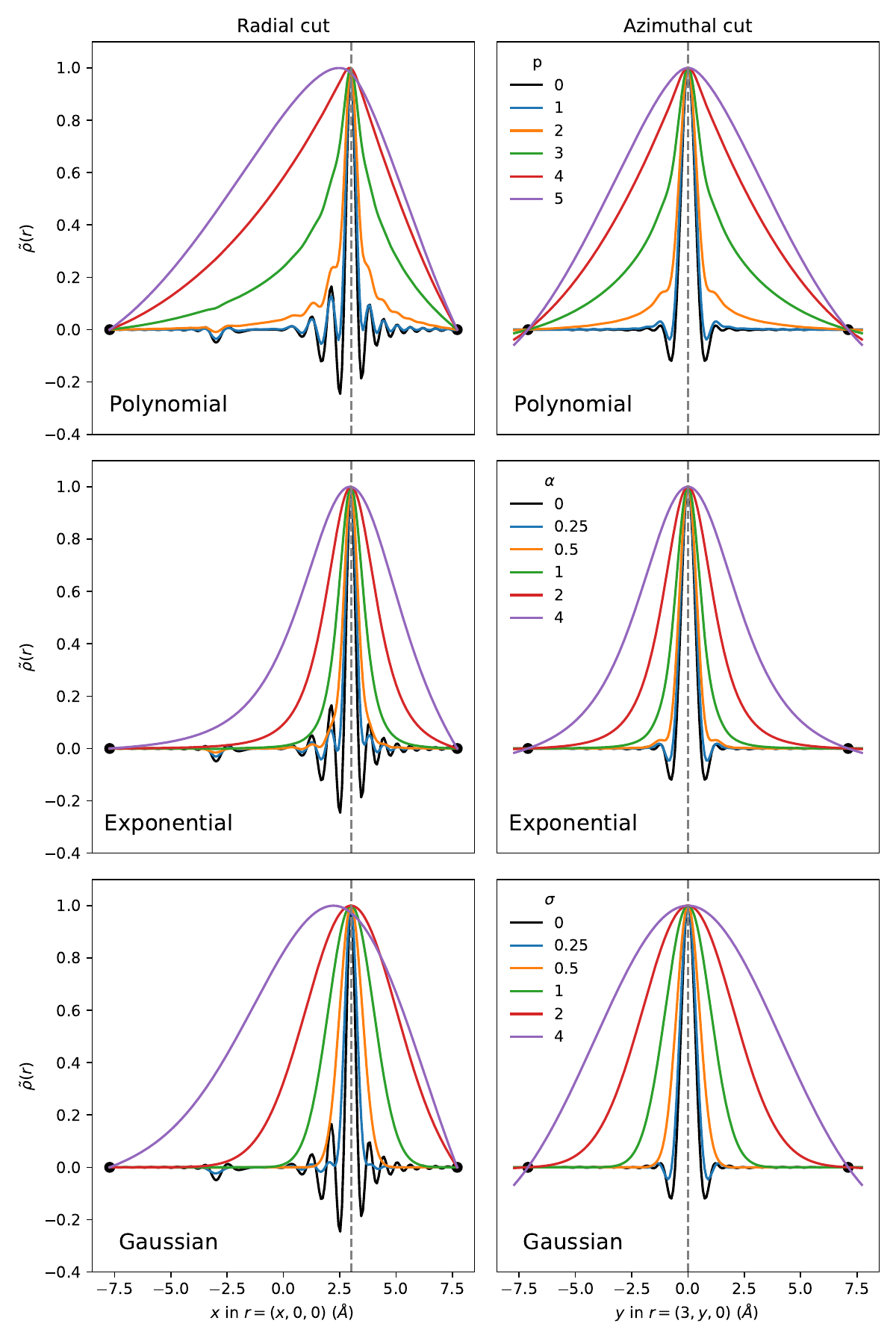}};
        
        \begin{scope}[x={(img.south east)}, y={(img.north west)}]
            \node[font=\small] at (0.49, 0.94) {(a)};
            \node[font=\small] at (0.95, 0.94) {(b)};
            \node[font=\small] at (0.49, 0.63) {(c)};
            \node[font=\small] at (0.95, 0.63) {(d)};
            \node[font=\small] at (0.49, 0.32) {(e)};
            \node[font=\small] at (0.95, 0.32) {(f)};
        \end{scope}
    \end{tikzpicture}
    \caption{The neighbor density $\tilde{\rho}$ corresponding to different choices of regularity prior are shown for a single neighbor atom at r=(3,0,0). The left-hand column shows a radial cut along $(r, 0, 0)$ while the right-hand column shows a perpendicular cut along $(3, r, 0)$. The black line shows the density with no regularity prior (a delta function represented using $n_\mathrm{max}=l_\mathrm{max}=20$) whilst the legend indicates the regularity parameter in $\lambda_{nl}^{-p/2}$, $\exp{\left(-\alpha\sqrt{\lambda_{nl}}\right)}$ and $\exp\left(-\sigma^2/2 \lambda_{nl} \right)$ for the polynomial, exponential and Gaussian priors respectively, where $\lambda_{nl}$ are the eigenvalues of $\Delta$. To aid visual comparison the maximum value of $\tilde{\rho}$ has been scaled to 1 for all curves. The black circles indicate the cutoff radius. The oscillatory behavior of the densities for low smoothing occur naturally in high-accuracy approximations of a delta function (cf. Dirichlet kernel and analogy with Methfessel-Paxton smearing \cite{methfessel1989high}) and may be beneficial in the context of very dense datasets.}
    \label{fig:density}
\end{figure}

For $p=2$ this is Poisson's equation, so that for $p$ even $\tilde{\rho}$ can be written as $p/2$ convolutions of the associated Green's function $G$ with the initial density $\rho_i^{(0)}$
\[
    \underset{\text{$p/2$ times}}{\underbrace{
        G \ast \cdots \ast G \ast 
    }}    \,  
    \rho_i^{(0)}.
\]
Each successive convolution leads to further broadening of the density, with $\phi \sim r^{p-3}$ for $p>0$ as shown in the top row of figure \ref{fig:density}.

One way to explore the limit as $p \to \infty$ is to consider regularity operators that decay super-algebraically. A seemingly natural choice is an exponential, 
$f_\tau(\lambda_{nl}) = \exp(- \tau \lambda_{nl})$,
or equivalently 
\begin{align*}
    f_\tau(-\Delta) = \exp(\tau \Delta).
\end{align*}
This choice is interesting if we note that the formal solution to the heat equation $\partial_tu = \Delta u $ can be written as $u(x,t) = \exp\left({t \Delta }\right)u(x,0)$, which then corresponds directly to $\tilde{\rho}_i =   \exp(\tau \Delta) \rho_i^{(0)}$. The heat equation can also be solved by convoluting the initial density with the heat kernel \cite{evans2022partial} which results in
\[
\tilde{\rho}_i = \Phi(\tau) \ast \rho_i^{(0)}.
\]
where, ignoring the previous boundary condition $\phi_{ nlm}(|\rr|=r_\mathrm{cut}) = 0$, $\Phi(\tau)$ is a normalized Gaussian of width $\sigma=\sqrt{\tau/2}$. 
As such, we see that applying the Gaussian regularity prior introduced in equation \eqref{eq:gauss_prior} $f_\tau(\lambda_{nl}) = \exp(- \tau \lambda_{nl}) \sim \exp(-\frac{1}{2}\sigma^2 (n+l)^2)$ approximately corresponds to forming ACE features from a SOAP-like Gaussian broadened neighbor density. This connection can also be made in a more explicit fashion, see Appendix \ref{sec:SOAP_explicit}, by directly expanding a Gaussian neighbour density in terms of eigenfunctions of the Laplacian and then extracting $f_\sigma(\lambda_{nl})$ from the expansion coefficients.

Finally, the effect of the exponential prior on the neighbour density can be understood in a similar manner by noting that $f(\lambda_{nl}) = \exp(-\alpha \sqrt\lambda_{nl})$ corresponds to a fractional heat equation,
\[
\tilde{\rho}_i  =  \exp( - \alpha (-\Delta)^{1/2} )  \rho_i^{(0)}.
\]
Interestingly this heat equation has an analytic kernel 
\[
H_\alpha = \frac{\alpha}{\pi^2(r^2 + \alpha^2)^2}, 
\]
which implies the associated neighbor density $\tilde{\rho} = H_\alpha \ast \rho_i^{(0)}$ has a Lorentzian shape. A comparison between the smoothing effects of all three priors, i.e. polynomial, exponential and Gaussian,  on $\tilde{\rho}$ is shown in figure \ref{fig:density} for $n_\mathrm{max}=l_\mathrm{max}=20$. Interestingly, despite the differences between the forms of the priors, the actual neighbor densities are strikingly similar once a moderate level of smoothing is applied. 

\subsection{Potential Fitting}

The \texttt{ACEPotentials.jl} package \cite{witt2023acepotentials} was used to fit all ACE models shown in the subsequent section. This package uses a radial basis of the form 

\begin{equation} \label{eq:ACEpot_radials}
    \begin{aligned}
        R_n(r_{ij}, Z_i, Z_j) &= R_{\zeta n'}(r_{ij}, Z_i, Z_j) \\
        &= \delta_{\zeta Z_j} P_{n'}(y_{ij})
        f_\mathrm{env}(r_{ij}, y_{ij}), \\  
            y_{ij} &= y(Z_i, Z_j, r_{ij}),
    \end{aligned}
\end{equation}
where $y_{ij} = y(Z_i, Z_j, r_{ij})$ denotes a coordinate transform, $f_{\rm env}$ is an envelope function enforcing predefined behavior as $r \to 0$ and $r \to r_{\rm cut}$, $P_n$ are shifted and scaled Legendre polynomials, while the chemical element dependence is folded into the composite index $n=(n', \zeta)$. Explicit functional forms are provided in~\cite{witt2023acepotentials}.
The regularity priors are therefore to be understood as acting in the transformed $y$-space and {\em after} factoring out the envelope function. The priors were applied using the definitions in \eqref{eq:decay_Cp_analytic} and \eqref{eq:gauss_prior} with $\sigma_{\mathrm{l}}=1.5\sigma_{\mathrm{n}}$ used for both the exponential and algebraic priors,  since this corresponds to the default $w_L$ parameter for determining basis set truncation~\cite{witt2023acepotentials}. For the Gaussian prior a different ratio, $\sigma_{\mathrm{n}}=\sqrt{5}\sigma_{\mathrm{l}}$, was used;  determined by the expansion of a Gaussian in the \ac{LE} basis of ref. \cite{bigi2022smooth}, see section \ref{sec:SOAP_explicit} for details. To better facilitate comparison to typical value used with \ac{SOAP} descriptors, all values of $\sigma$ for the Gaussian prior will be quoted in \AA{} where, $\sigma_n=\sqrt{5} \sigma/r_\mathrm{cut}$ and $\sigma_l= \sigma/r_\mathrm{cut}$. 

To promote repulsion at very short inter-atomic distances, where Coulomb repulsion between nuclei dominates \cite{ziegler1985stopping, nordlund2025repulsive}, the training data was augmented with a single synthetic dimer configuration labeled with energy $E_\mathrm{syn}$=100 eV and with separation $r_\mathrm{syn}$, defined by $f_\mathrm{env} (r_\mathrm{syn}) = E_\mathrm{syn}$ for each pair of elements. Observations of energies, forces and stresses were given relative weights of $w_E=30$, $w_F=1$ and $w_S=1$ respectively while much smaller weights of $w_\mathrm{syn}=0.01$ used for the synthetic dimer data. Furthermore, a purified version of the ACE basis was used, where the two-body contributions were removed from all higher order terms, see refs. \cite{witt2023acepotentials, ho2024atomic} for a detailed discussion.

\section{Results}
Having examined different possible forms for the regularity prior and how they can be interpreted, we now present a diverse set of numerical tests devised to asses the effect of the priors in realistic use cases. We include a combination of quick metrics, such as pointwise error estimates and 1D cuts through the PES, and more involved tests which probe stability and phase transitions during \ac{MD} simulations. Silicon is used as the primary test system with additional experiments carried out on the aspirin molecule to ensure the results are not specific to single-element or condensed phase systems.

\subsection{Silicon}
The Silicon-GAP-18 dataset from \cite{bartok2018machine} was used to investigate the practical consequences of fitting ACE potentials using various regularity priors. This database is very structurally diverse, containing a mixture of crystalline, amorphous and liquid configurations as well as various crystalline surfaces, defects and dislocations. This diversity ensures good coverage of the silicon potential energy surface and made it possible o fit a general-purpose potential \cite{bartok2018machine} using the \ac{GAP} \cite{bartok2010gaussian, deringer2021gaussian, klawohn2023gaussian} framework.
Specifically we fit ACE potentials to the original dataset and a smaller subset, referred to as Si10pc, containing $\sim$10\% of the original data. 
We focus on results obtained on Si10pc, where the differences between different choices of priors should be more pronounced. Recent work has shown that a carefully selected 5\% of the original dataset is sufficient to fit high-quality ACE potentials \cite{van2023hyperactive}; by contrast, Si10pc was constructed by applying a random 10:90 train:test split to the full dataset where the split was stratified by both configuration size and type. Additionally, the most and least dense configuration of each type was included in the training set to avoid excessive extrapolation in  density. 

\begin{figure}[htbp]
    \centering
    \begin{tikzpicture}
        \node[anchor=south west, inner sep=0] (img) 
            {\includegraphics[width=0.5\textwidth]{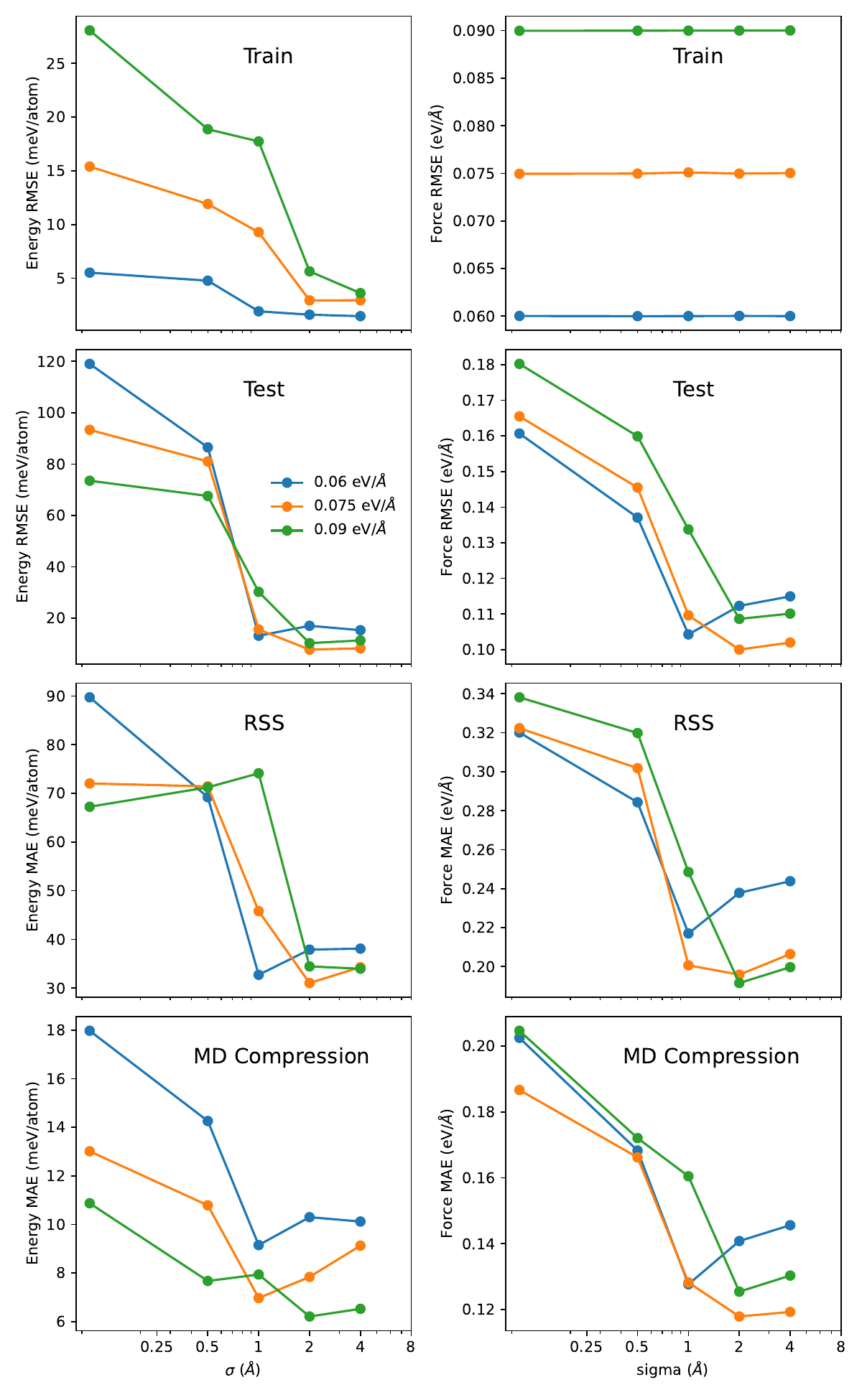}};
        
        \begin{scope}[x={(img.south east)}, y={(img.north west)}]
            \node[font=\small] at (0.45, 0.97) {(a)};
            \node[font=\small] at (0.95, 0.97) {(b)};
            \node[font=\small] at (0.45, 0.73) {(c)};
            \node[font=\small] at (0.95, 0.73) {(d)};
            \node[font=\small] at (0.45, 0.49) {(e)};
            \node[font=\small] at (0.95, 0.49) {(f)};
            \node[font=\small] at (0.45, 0.07) {(g)};
            \node[font=\small] at (0.95, 0.07) {(h)};
        \end{scope}
    \end{tikzpicture}
    \caption{Force and energy errors are shown for ACE models fit using Gaussian regularity priors on configurations in the training and test sets, as well as two different configuration types from ref. \cite{morrow2022indirect} that are not present in the training data, see main text for details . Larger values of $\sigma$ correspond to a more regularized prior. The force error on the training set, shown in the legend, is constant as this was used to select the strength of the regularization.}
    \label{fig:10_90_gaussian_errors}
\end{figure}
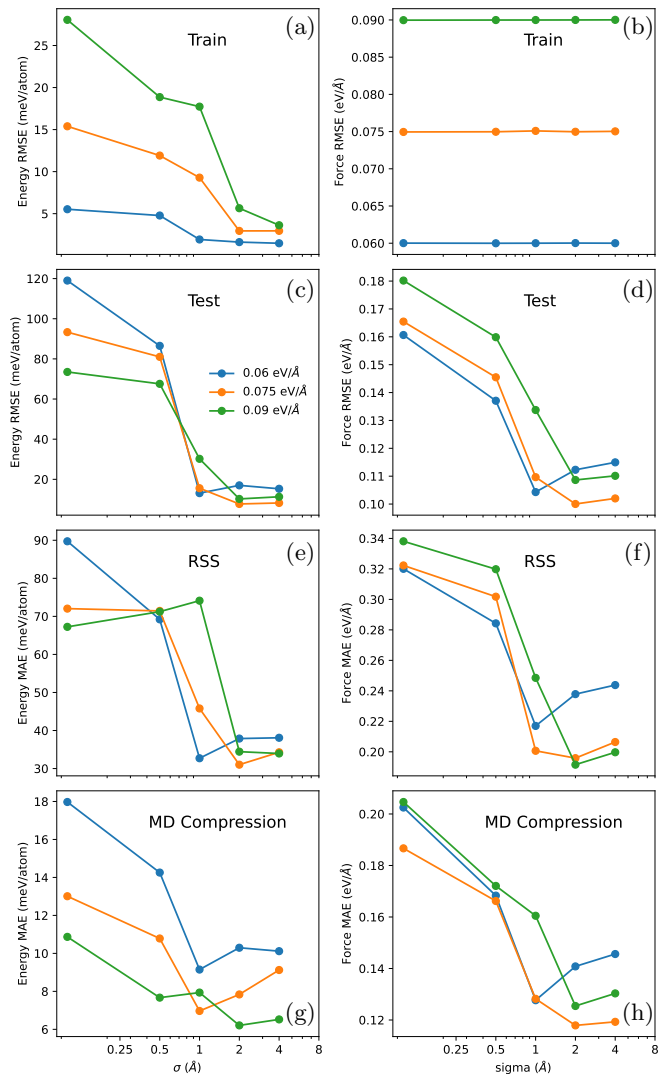

\begin{figure}[h!]
    \centering
    \includegraphics[width=0.45\textwidth]{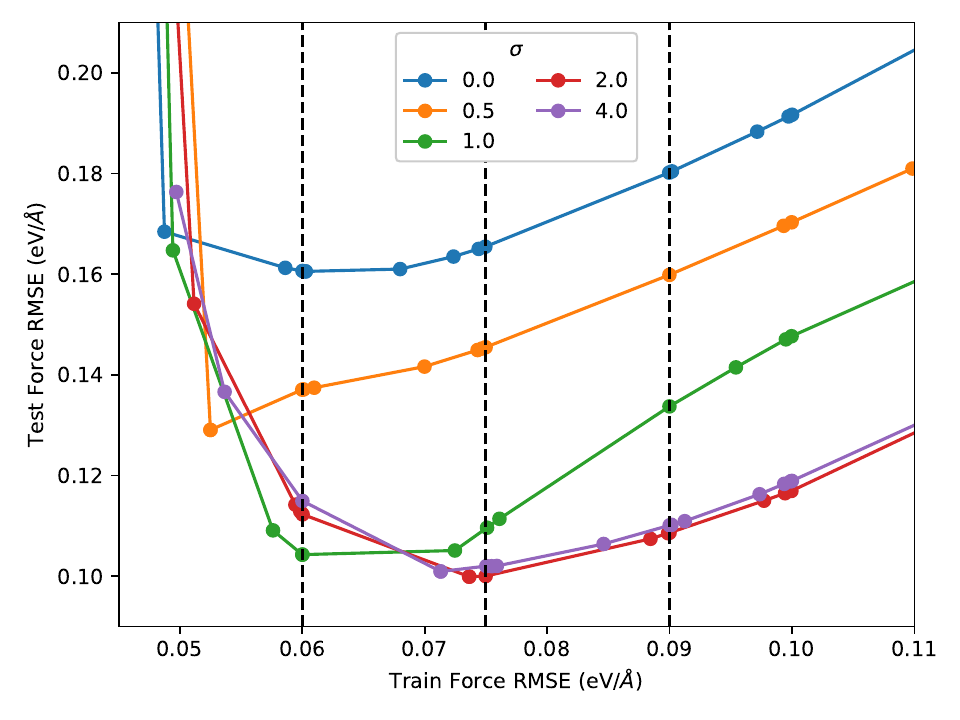}
    \caption{Force RMSE on the training and test sets is shown for potential fit to Si10pc using a Gaussian regularity prior with varying $\sigma$. Each marker corresponds to a particular choice of Tikhonov regularization strength. The dashed black lines correspond to the three different training force RMSEs shown in figure \ref{fig:10_90_gaussian_errors}.}
    \label{fig:10_90_FRMSE}
\end{figure}

The ACE potentials were constructed using a $r_{\rm cut} = 5~\text{\AA}$ local cutoff and correlation orders up to $N_{\rm max}=4$ (5-body) features, with a maximum polynomial degree of 21 for all correlation orders. The potentials were then fit using simple Tikhonov regularization applied after the regularity prior was used to scale the basis functions appropriately as outlined in Section \ref{sec:rscaling}. To distinguish between the strength and the ``shape'' of the regularization, i.e. different form of regularity prior, the regularization strength was tuned such that the force root mean square error (RMSE) on the training set was equal for all models. This was repeated for three different target training force RMSEs with the resulting energy and force errors on the training and test sets shown in figure \ref{fig:10_90_gaussian_errors} for models fit using the Gaussian regularity prior. Errors were also evaluated for two further configurations types: 100 configurations from \ac{RSS}\cite{pickard2011ab} and 100 configurations from a \ac{MD} simulation where amorphous Si was compressed from 0 to 20 GPa. These configurations were taken from ref \cite{morrow2022indirect} and are used to illustrate the effect of regularity priors in a highly extrapolative regime. For all configuration types, using $\sigma=2$ \AA{}
lead to $\sim$ 40\% reduction in force RMSE and up to 80\% reduction in energy RMSE compared to $\sigma=0$ (no regularity prior). 
%As shown in figure \ref{fig:10_90_FRMSE}, this is not caused by the specific choice of target force RMSE, with these three values covering the minimum achievable test force RMSE for all values of $\sigma$.
As is evident from figure 4, which shows the error as a function of $\sigma$  for three different choices that cover the minimum achievable test force RMSE, this is achieved regardless of the specific target force RMSE
It is notable that despite the equivalence between the Gaussian regularity priors and SOAP features, outlined in Section \ref{sec:rscaling}, the optimal value for $\sigma$ is much larger than would typically be used in SOAP-GAP potentials where $\sigma=0.3-0.5$ \AA{} is more typical \cite{deringer2018data}. This can be partially attributed to the distance transform applied to the radial coordinate, which increases the resolution near the typical nearest-neighbor position, effectively reducing $\sigma$ where there is most data and increasing it elsewhere. We also note that the exact shape of the regularity prior appears to make little difference to the achievable errors, with very similar performance achieved using both the algebraic and exponential regularity priors as shown in figure \ref{fig:10_90_prior_comp}. Given this, we only present results for the Gaussian prior (Force RMSE on the training set = 0.075 eV/\AA{}) in the main text, with corresponding Figures for the algebraic and exponential priors provided in figures \ref{fig:10_90_algebraic} and \ref{fig:10_90_exponential} in the Appendix. 

\begin{figure}[h]
    \centering
    \begin{tikzpicture}
        \node[anchor=south west, inner sep=0] (img) 
            {\includegraphics[width=0.5\textwidth]{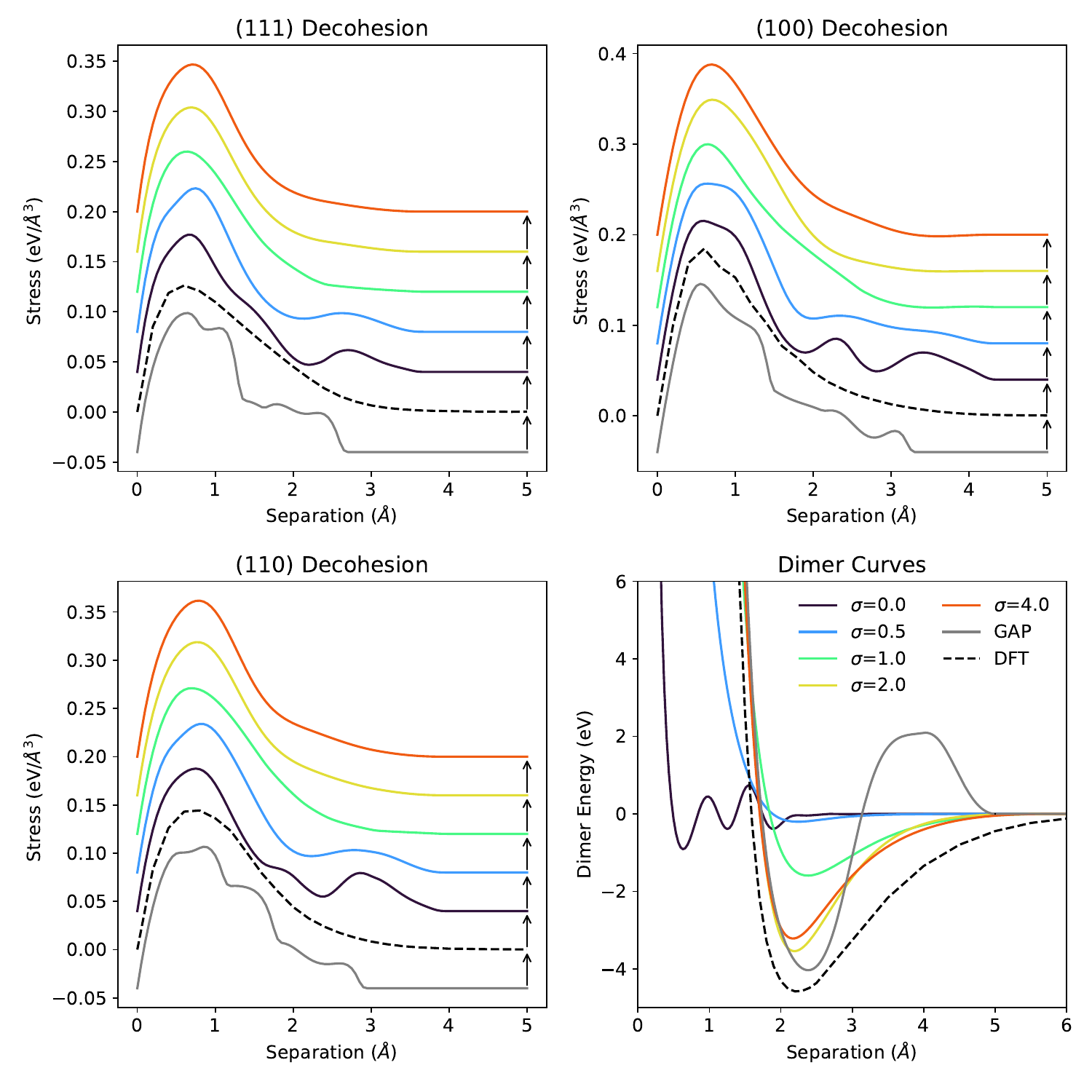}};
        
        \begin{scope}[x={(img.south east)}, y={(img.north west)}]
            \node[font=\small] at (0.47, 0.93) {(a)};
            \node[font=\small] at (0.95, 0.93) {(b)};
            \node[font=\small] at (0.47, 0.43) {(c)};
            \node[font=\small] at (0.95, 0.1) {(d)};
        \end{scope}
    \end{tikzpicture}
    \caption{(a)-(c))The stress (derivative of energy) is plotted as a function of separation for decohesion of silicon diamond into the (100), (110) and (111) surfaces.  The curves for different values of $\sigma$ are offset (black arrows) to aid comparison. (d) The bottom right panel shows the energy of the Si-Si dimer as a function of separation between the atoms.}
    \label{fig:Si_10_90_gaussian_decohesion}
\end{figure}
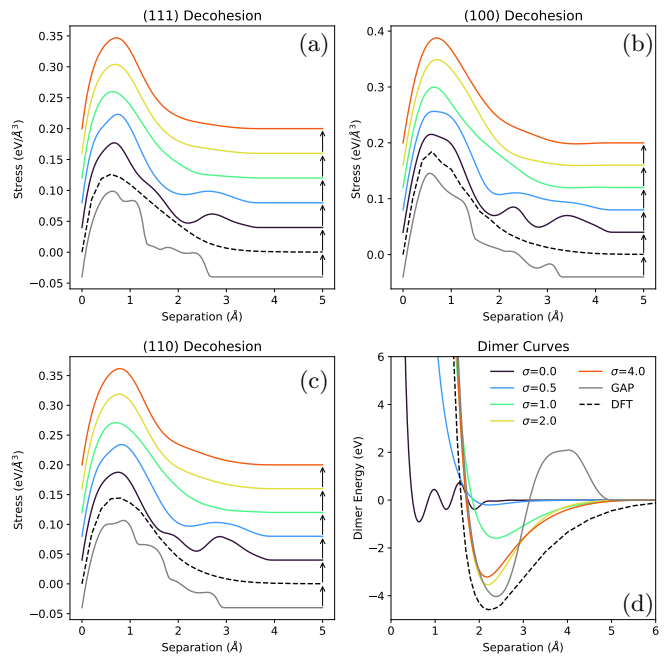

Whilst reducing errors is highly desirable it is not the only important metric. For instance,  many modern architectures achieve excellent errors but are unable to perform stable \ac{MD} simulations \cite{fu2022forces}, or do so at the cost of having very rough potential energy surfaces with many spurious additional minima, see figure \ref{fig:Mg_dimers}. Quantifying this is challenging but we can probe it by simply plotting various cuts through the potential energy surface. The simplest example of this is the Si-Si dimer interaction energy which is shown in the bottom right of figure \ref{fig:Si_10_90_gaussian_decohesion}. Despite all models having the same training force RMSE, the models with large values of $\sigma$ bear much closer resemblance to the reference \ac{DFT} dimer curve. Crucially,  they exhibit no false minima and are repulsive at close-approach, despite there being no dimer data in the training set.

\begin{figure}[h]
    \centering
    \begin{tikzpicture}
        \node[anchor=south west, inner sep=0] (img) 
            {\includegraphics[width=0.5\textwidth]{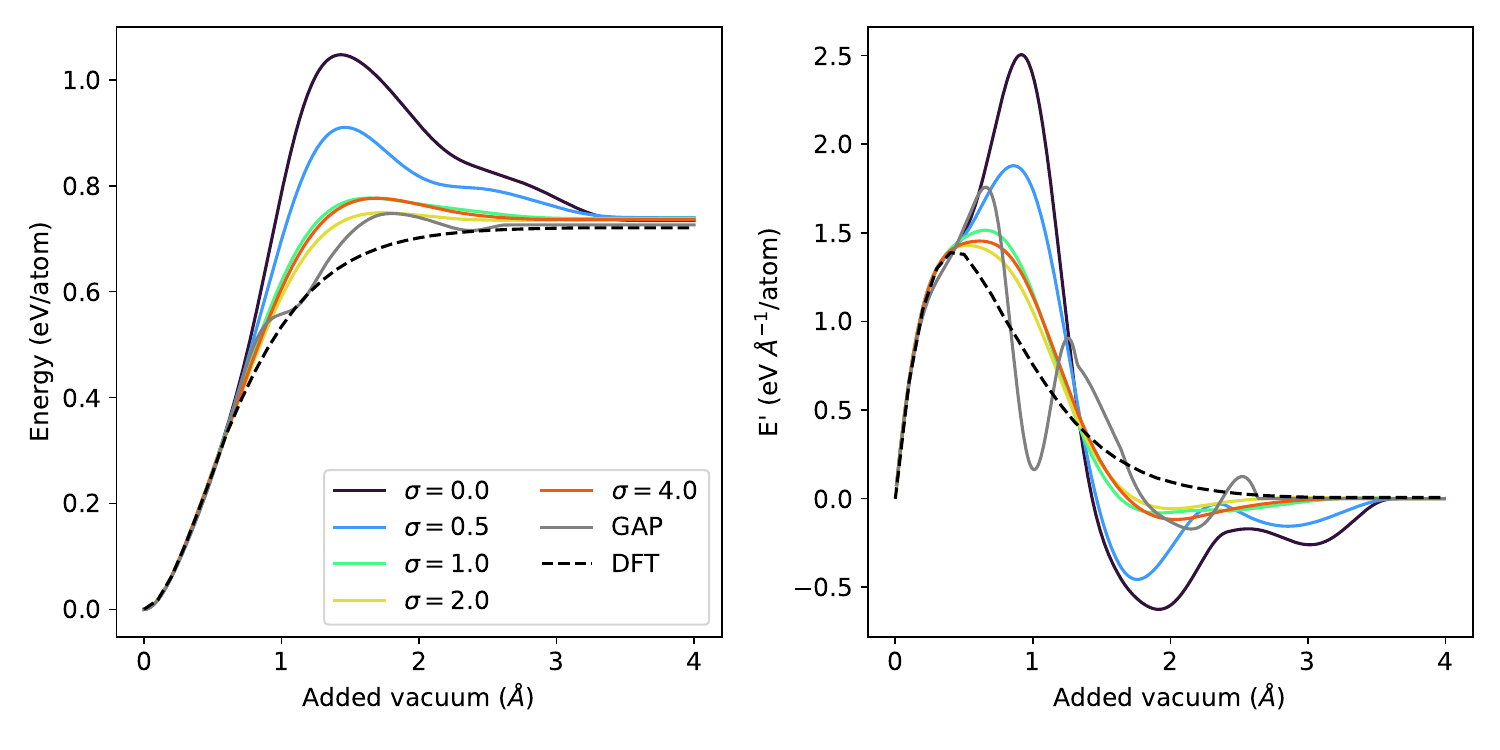}};
        \begin{scope}[x={(img.south east)}, y={(img.north west)}]
            \node[font=\small] at (0.11, 0.9) {(a)};
            \node[font=\small] at (0.61, 0.9) {(b)};
        \end{scope}
    \end{tikzpicture}
    \caption{The energy (a) and its derivative with respect to added vacuum (b) are shown as the diamond structure silicon crystal is rigidly exfoliated into layers of silicene.}
    \label{fig:silicene}
\end{figure}

Inspecting higher body-order terms individually is both impractical and challenging to visualize due to the increased number of parameters. As such, we instead  analyze them by constructing a series of 1D cuts through the full PES that depend on all body-orders present in the potential. The first example was taken from ref. \cite{bartok2018machine} and involves computing the decohesion energy of the Si-diamond structure as it is rigidly separated to create a slab with two surfaces separated by vacuum. The derivative of the decohesion energy, which has the units of stress, is shown in figure \ref{fig:Si_10_90_gaussian_decohesion} for trajectories that created (100), (110) and (111) surfaces . The reference DFT stress is notably smooth, with a single maximum and no further oscillations. In contrast, the $\sigma=0$ ACE potentials show clear spurious oscillations which are gradually damped away as $\sigma$  is increased. Smaller differences are observed in the decohesion energy itself, shown in figure \ref{fig:10_90_gaussian_decohesion_E}, with variations in $\sigma$ having little effect on the final surface energy; this is expected as these configurations are present in the training set. A similar test was performed whereby Si-diamond was rigidly exfoliated to individual layers of silicene. Here, clear differences in the energy are observed between the potentials with all ACE models having an additional energy barrier to exfoliation that is not present in the \ac{DFT} reference; see the left panel of figure \ref{fig:silicene}. The height of this barrier, and the size of associated oscillations in the derivative of the energy, decrease as $\sigma$ is increased.

\begin{figure}[h]
    \centering
    \begin{tikzpicture}
        \node[anchor=south west, inner sep=0] (img) 
            {    \includegraphics[width=0.5\textwidth, clip=true, trim={0 10cm 0 10.9cm}]{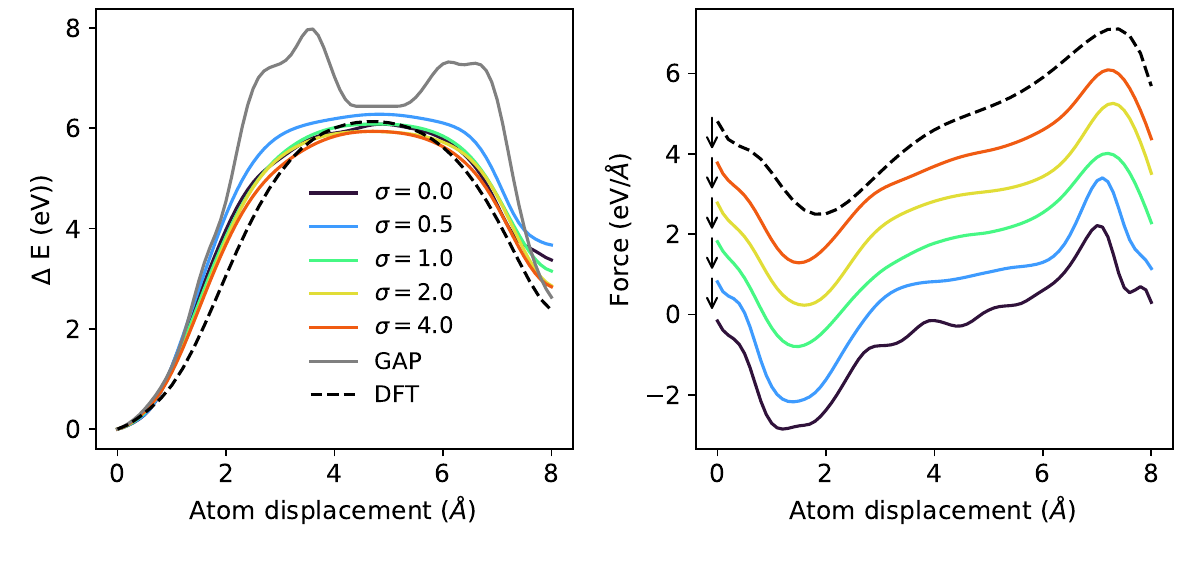}};
        \begin{scope}[x={(img.south east)}, y={(img.north west)}]
            \node[font=\small] at (0.11, 0.91) {(a)};
            \node[font=\small] at (0.61, 0.91) {(b)};
        \end{scope}
    \end{tikzpicture}
    \caption{(a) The change in energy as an atom is move across the vacuum from the bottom (110) surface of a silicon  slab to the top (110) surface. (b) The z component of the force on the atom being moved is shown. Curves for different values of $\sigma$ are offset (black arrows) to aid comparison.}
    \label{fig:rigid_surface_to_surface}
\end{figure}

In the next test an atom in the top layer of a Si slab, a (110) surface, was moved in the $z$ direction until the atom sat just below the bottom surface of the slab due to periodic boundary conditions. The energy and z-component of the force on the target atom are shown in figure \ref{fig:rigid_surface_to_surface}; as before, increasing the regularity prior leading to better agreement with the \ac{DFT} reference, particularly for the force. Notably, using larger $\sigma$ smooths out the unphysical oscillations present from 2 \AA{} and onwards but does not smooth out the relatively sharp feature around 0.5 \AA{} which is present in the \ac{DFT} reference. The same test was repeated for the (100) and (111) surfaces with the results shown in figure \ref{fig:10_90_gaussian}.  

%{Figures/Silicon/10_90_split/10_90_gaussian_volume_scans_combined.pdf}
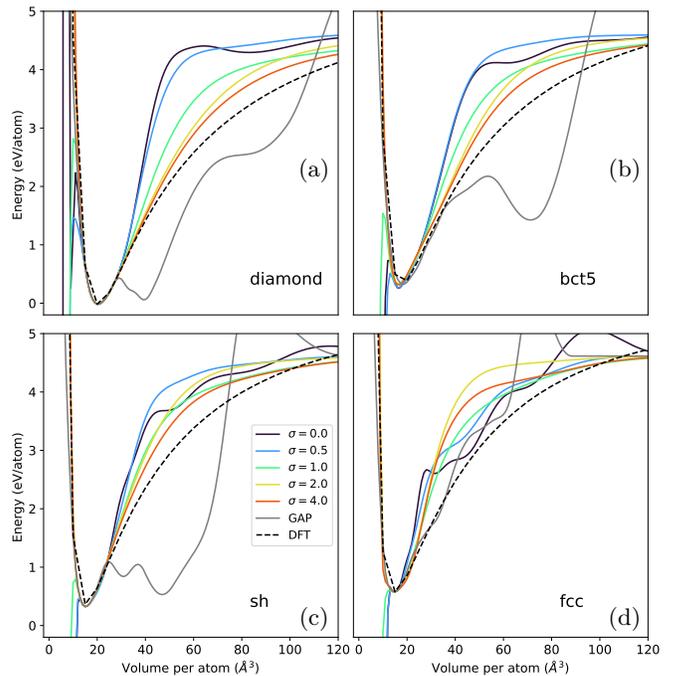
\begin{figure}[h]
    \centering
    \begin{tikzpicture}
        \node[anchor=south west, inner sep=0] (img) 
            {    \includegraphics[width=0.5\textwidth,  trim={0cm 0cm 0cm 22cm}, clip]{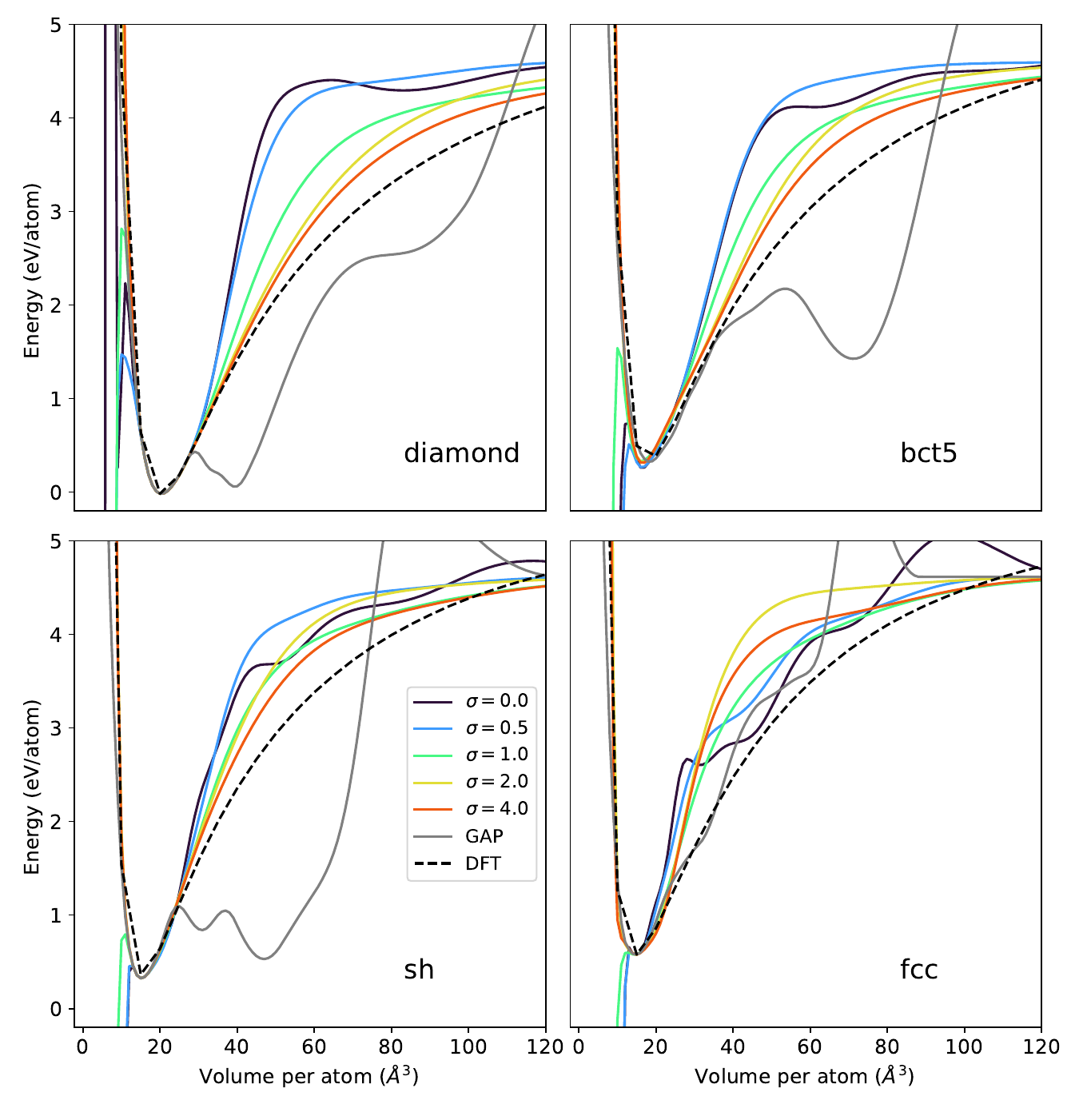}};
        \begin{scope}[x={(img.south east)}, y={(img.north west)}]
            \node[font=\small] at (0.47, 0.75) {(a)};
            \node[font=\small] at (0.93, 0.75) {(b)};
            \node[font=\small] at (0.47, 0.11) {(c)};
            \node[font=\small] at (0.93, 0.11) {(d)};
        \end{scope}
    \end{tikzpicture}
    \caption{The energy of various silicon crystal structures is shown as the volume is varied with fractional positions of the atoms held constant. The corresponding pressure-volume curves (derivatives) are provided in Fig. \ref{fig:10_90_gaussian}. }
    \label{fig:volume_scans}
\end{figure}

As a final test, the energy of various know Si phases, including diamond, bct5 and the simple-hexagonal (sh) polymorph, were computed as the volumes were varied isotropically. 
The resulting energy  vs volume curves are provided in figure \ref{fig:volume_scans} with pressure vs volume in figure \ref{fig:10_90_gaussian}. Under compression, holes are present in the potentials with $\sigma=0$, 0.5 and 1.0~\AA{}, despite the dimer curves for $\sigma=0.5~\AA{}$ and $\sigma=1.0~\AA{}$ being repulsive at close-approach. Increasing $\sigma$ removes this issue and is a clear sign that the regularity prior is improving the behavior (and extrapolation) of the higher body-order terms. All ACE potentials correctly reproduce the curvature near the equilibrium volume whilst those with larger regularity priors capture more of the an-harmonic behavior at larger volumes. It is worth noting that the GAP and $\sigma=0$ potentials both exhibit additional minima in the energy-volume curves at large volumes. The large energy barriers to these states means they would not be encountered during typical MD simulations initialized near equilibrium conditions. However, they could cause issues in e.g. random structure search (see figure \ref{fig:Si_RSS}) or nested sampling, which explore a much wider range of phase space. 

Next, ACE potentials were fit to the full Silicon-GAP-18 dataset. The aforementioned tests were repeated with the new models, see figure \ref{fig:full_dataset_gaussian}, and similar improvements were seen when utilising a regularity prior, suggesting that such priors still improve performance even in a much more data-rich setting. Secondly, compression simulations were performed, wherein the low-density amorphous (LDA) structure of Si from \cite{deringer2021origins} was gradually compressed during a \ac{MD} simulation by applying a pressure ramp from zero to 20 GPa. It is expected \cite{deb2001pressure, mcmillan2005density, pandey2011pressure} that the material undergoes a sequence of phase transitions: LDA $\rightarrow$ (very high density amorphous) VHDA  $\rightarrow$ polycrystalline simple-hexagonal (pc-sh). This sequence is known to be reproduced by GAP models trained on the full Silicon-GAP-18 dataset \cite{deringer2021origins}, and can serve as a challenging test for interatomic potentials \cite{morrow2022indirect, morrow2023validate}.
Indeed, it was previously shown that the nucleation of pc-sh was not consistently produced by Moment Tensor Potentials (MTPs) trained directly on the Si-GAP-18 dataset \cite{morrow2022indirect}. Instead, consistent nucleation could only be achieved with MTP models if the training dataset was augmented with large, 1000 atom, configurations spanning the desired pressure range but labeled with a GAP model fit to the original data. This ``student--teacher'' model relationship was exploited to transfer the desired behaviour from the accurate, but slower, GAP model to the much faster MTP, enabling larger simulations to be performed. In the present work, we explore whether using a regularity prior might allow us to bypass the need for the addition of synthetic data and instead to achieve the desired sequence of phase transitions with ACE models fitted directly to the Silicon-GAP-18 dataset. 

\begin{figure*}
	\centering
        \begin{tikzpicture}
        \node[anchor=south west, inner sep=0] (img) 
            {\includegraphics[width=1.0\textwidth,  clip=true, trim={0 7cm 0.5cm 0}]{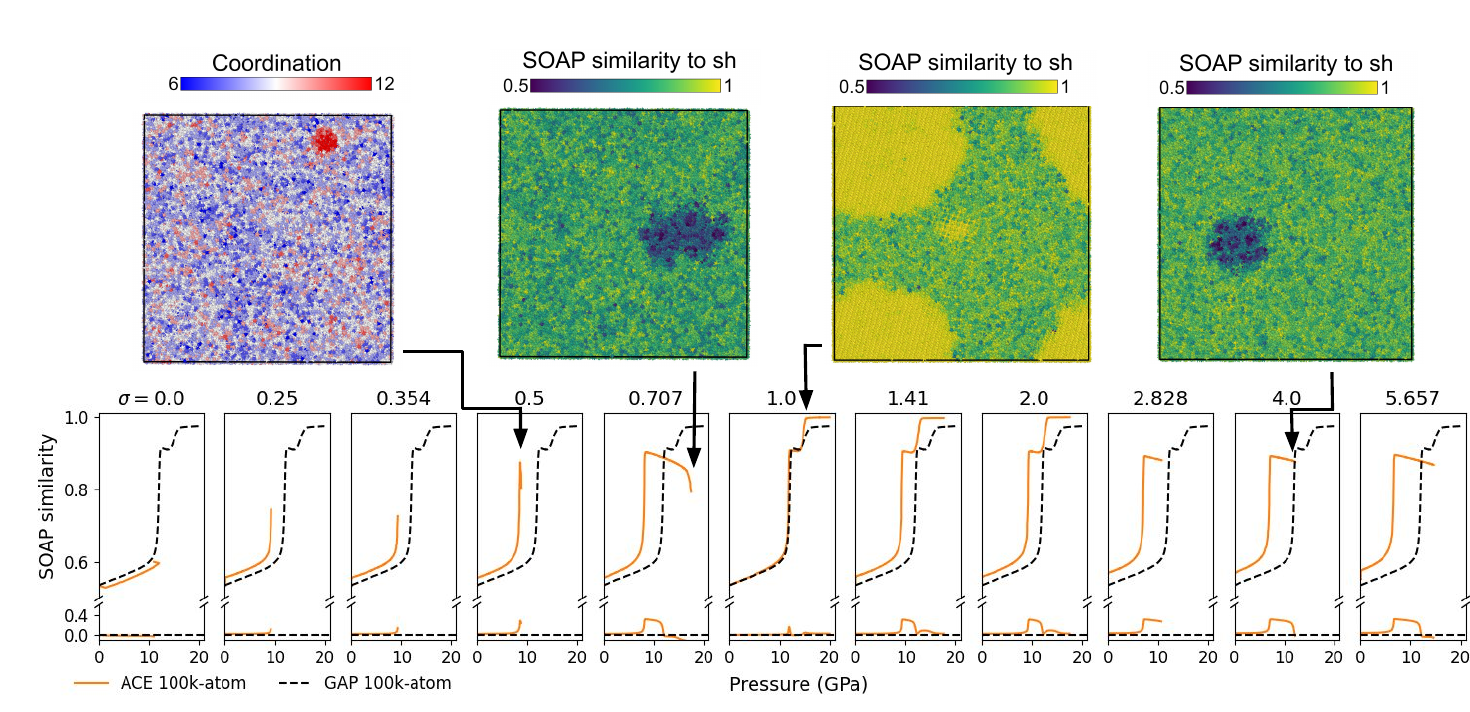}};
        \begin{scope}[x={(img.south east)}, y={(img.north west)}]
            \node[font=\small] at (0.09, 0.91) {(a)};
            \node[font=\small] at (0.33, 0.91) {(b)};
            \node[font=\small] at (0.56, 0.91) {(c)};
            \node[font=\small] at (0.78, 0.91) {(d)};
            \node[font=\small] at (0.04, 0.45) {(e)};
        \end{scope}
    \end{tikzpicture}
	\caption{(e) Following ref. \cite{morrow2022indirect}, the average SOAP similarity to the pressure adjusted sh-Si phase is shown for 100,000 atom simulations of amorphous Si during compression from 0-20 GPa. The sharp increase around 10 GPa indicates the LDA $\rightarrow$ VHDA transition whilst the subsequent increase occurs on nucleation of the pc-sh phase. The difference in SOAP similarity compared to the GAP reference is shown below (note the break in the y-axis) whilst representative snapshots are highlighted above. (a) Atoms are colored according to local coordination number to highlight an erroneously highly coordinated cluster in the top right corner which occurs due to a hole in the potential. (b) and (d) No nucleation of the sh-Si phase is seen with some regions adopting an unusual local structure with hints of hexagonal symmetry (c) nucleation of the sh-crystal is seen with multiple differnet nucleation sites present. Similar structures are seen with $\sigma=1.41$ and 2.0~\AA{}, with simulations remaining stable as the growing grains impinge on each other, likely inducing large local stresses. }
	\label{fig:compression_MD}
\end{figure*}
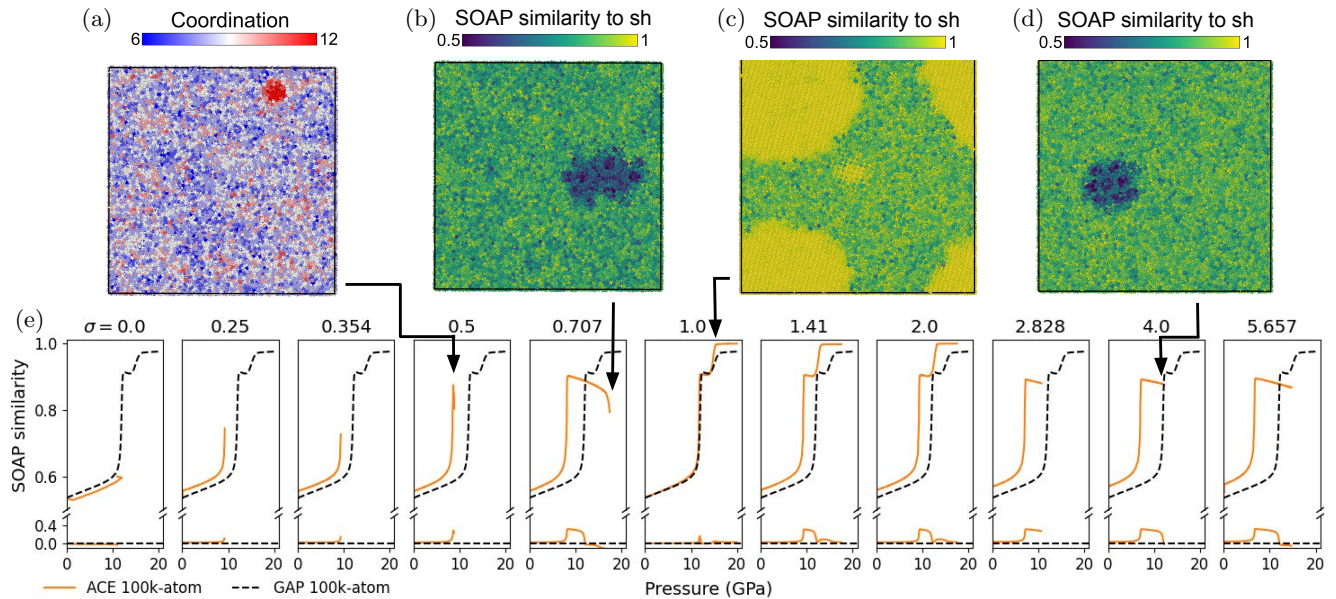

\begin{figure*}
    \centering
    \begin{tikzpicture}
        \node[anchor=south west, inner sep=0] (img) 
            {\includegraphics[width=0.95\textwidth]{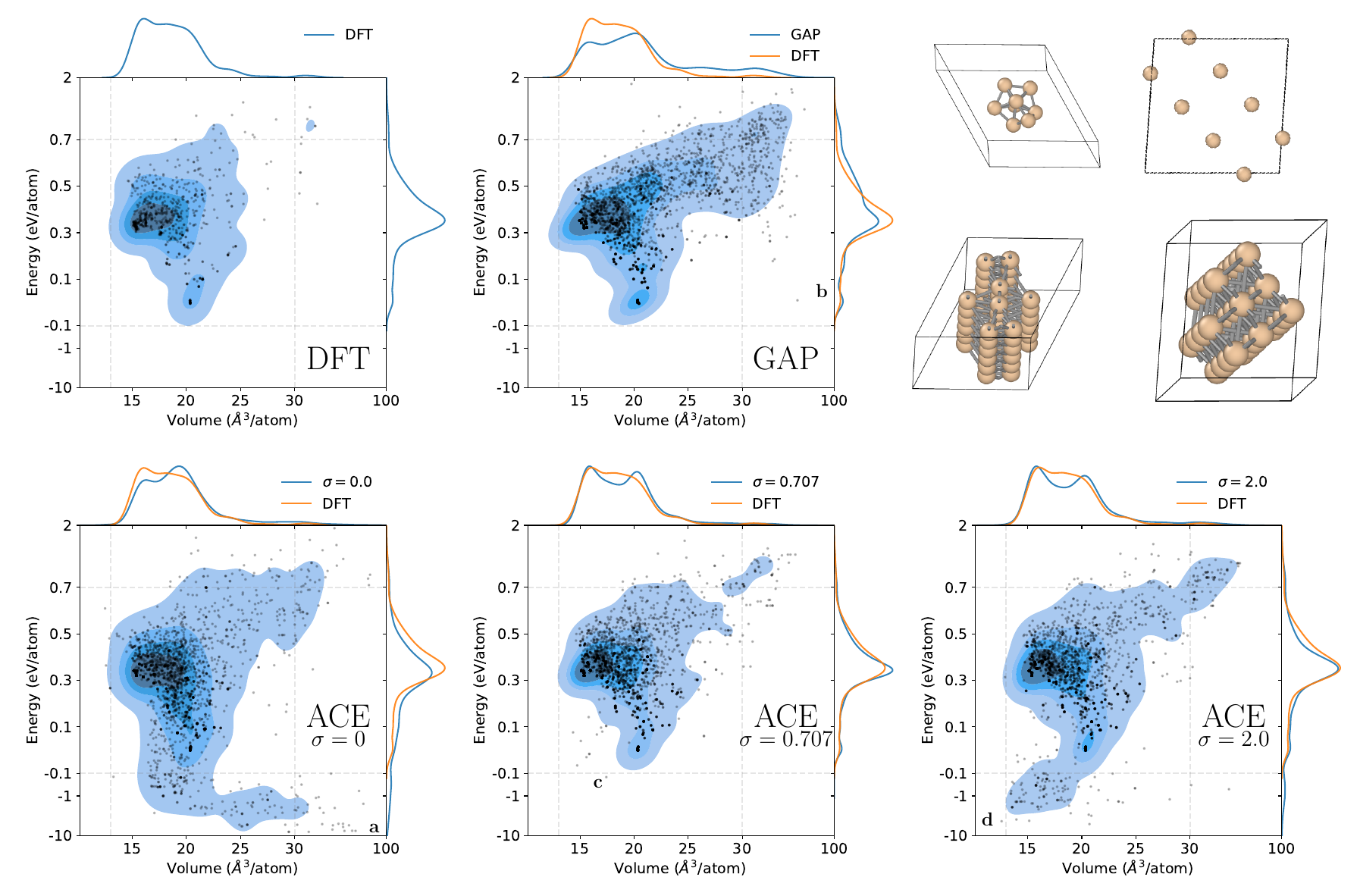}};
        \begin{scope}[x={(img.south east)}, y={(img.north west)}]
            \node[font=\small] at (0.05, 0.95) {(a)};
            \node[font=\small] at (0.39, 0.95) {(b)};
            \node[font=\small] at (0.67, 0.95) {(c)};
            \node[font=\small] at (0.82, 0.95) {(d)};
            \node[font=\small] at (0.67, 0.71) {(e)};
            \node[font=\small] at (0.82, 0.71) {(f)};
            \node[font=\small] at (0.05, 0.45) {(g)};
            \node[font=\small] at (0.39, 0.45) {(h)};
            \node[font=\small] at (0.73, 0.45) {(i)};
        \end{scope}
    \end{tikzpicture}
    \caption{The distribution of energy, relative to the diamond structure, and volume of random structures relaxed with various energy methods are shown. The axes are linear within the central square bounded by dashed grey lines and logarithmic outside. Four illustrative structures which are all predicted to be overly stable are show in (c) consists of isolated clusters, (d) is a 3D lattice where the closest distance between any pair of atoms is 3.7\AA~ and (e) and (f) are both 1D ``nano-wires'' where some atoms have more than 4 immediate neighbours.}
    \label{fig:Si_RSS}
\end{figure*}

The results of such compression simulations, carried out as in ref. \cite{morrow2022indirect},  are shown in figure \ref{fig:compression_MD}, where the SOAP similarity to the pressure adjusted pc-sh phase is shown as a function of pressure for ACE models fit with 11 different levels of Gaussian regularity prior. With the default prior, $\sigma=0$, the dynamics become unstable during the transition from LDA $\rightarrow$ VHDA. Specifically, the model predicts spuriously large forces that accelerate atoms to velocities exceeding the integrator's stability threshold, resulting in lost atoms and termination of the simulation. Prior to the crash, we observe the formation of chemically-unreasonable clusters of highly-coordinated Si, visible as blue patches in Figure\ \ref{fig:compression_MD}B–C. These are false minima (``holes'') in the PES accessed under the influence of compression. Increasing $\sigma$ leads to a gradual increase in stability, with the $\sigma=1.0$~\AA{} model successfully completing the full simulation. With moderate levels of regularity prior,  $\sigma=1.0-2.0$~\AA{}, nucleation of the pc-sh phase is successfully observed, indicating that a well chosen regularity prior has helped mitigate the need for the student-teacher relationship. At larger values of $\sigma$ the dynamics becomes less stable, with no successful nucleation found to occur. This worsening performance correlates with the increasing errors shown in figure \ref{fig:full_silicon_errors}, and implies that a  well chosen $\sigma$ is one that approximately minimizes the force error on the test set. 

Following this, the Silicon ACE potentials were used to perform a random structure search. The search was restricted to 8 atom unit cells with 1000 putative structures generated without symmetry constraints and a further 2000 generated using 2-4 randomly chosen symmetry operations. Minimum distance constraints of 1.7\AA~ were enforced between neighbouring atoms and the volume of the initial cell was drawn from a uniform distribution spanning 12-30 $\text{\AA}^3$/atom.  Structural relaxations were carried out with 0.01 GPa of external pressure applied in lieu of a dispersion-correction scheme and a force tolerance of 0.001 eV/$\text{\AA}$. 
%Reference \ac{DFT} relaxations were performed using the identical settings used to generate the original dataset whilst a slightly larger tolerances of 0.01 eV/$\AA$ and 0.01 GPa were used for efficiency. 
The distribution of relaxed structures is shown in figure \ref{fig:Si_RSS} for searches performed using \ac{DFT} (CASTEP \cite{clark2005first}), GAP and representative ACE models. The GAP search correctly identified the diamond structure as lowest in energy but contains a much larger proportion of low density, high energy structures than the \ac{DFT} reference. In contrast, the ACE models all incorrectly find at least one structure to be lower in energy than diamond, with such structures all comprised of locally dense, highly coordinated, Si clusters or nano-wires. Among the ACE potentials, using a non-zero $\sigma$ significantly improves the marginal energy and volume distributions compared to the \ac{DFT} reference and helps removes the low density, low energy structures. This result might have been expected from figure \ref{fig:volume_scans} where the spurious oscillations in the volume distribution are eliminated. However, we stress that a regularity prior will not remove all spurious minima in the predicted PES and indeed this is the case, as is seen for the search performed with $\sigma=2.0$~\AA{}. Perhaps the clearest benefit of applying the regularity prior is the fraction of structural relaxations that completed successfully. Structures were relaxed using the LBFGS algorithm \cite{liu1989limited} with a force tolerance of 0.05 eV/\AA{} for \ac{DFT} and 0.001 eV/\AA{} for all potentials. \ac{DFT} optimizations used a maximum of 200 steps, whilst with potentials a maximum of 20 batches of 100 steps was used with 0.001 \AA{} of Gaussian noise added to the positions every 100 steps to aid convergence.  With \ac{DFT} and GAP the number of failures was $<$1\% whilst with $\sigma=0$ \AA{} $\sim$ 11\% failed. For $\sigma=0.707$ \AA{} this dropped back below 1\% and remained below 2\% for all larger values. 

\subsection{Aspirin}
The effect of applying the Gaussian regularity prior was also tested on ACE potentials fit to Aspirin \ac{MD} trajectories from the rMD17 dataset \cite{christensen2020role}. This data was chosen as it relates to an organic molecule and contains multiple chemical element, both in contrast with the single-element condensed-phased Silicon dataset. In SOAP-GAP potentials it is customary to specify a different value of $\sigma$ for each element, with larger values used for larger elements. Here, $\gamma_{nl}$ has no element dependence as the scaling of the basis functions $\gamma_{nl}$ depends on $\sigma/r_\mathrm{cut}$, see Equation \ref{eq:exp_reg}, and we choose to scale both $\sigma$ and the cutoff $r_\mathrm{cut}$ in the same way for different element-pairs.  A convenient consequence of this is that a single value of $\sigma$, combined with the default scaling based on tabulated covalent radii, specifies the regularity prior completely. In practice,  $\gamma_{nl}$ is computed using the largest cutoff present for any element-pair, so the specified value of $\sigma$ corresponds to this cutoff and the effective $\sigma$ for all other element-pairs is reduced accordingly. 

\begin{figure}[htbp]
    \centering
    \includegraphics[width=0.5 \textwidth]{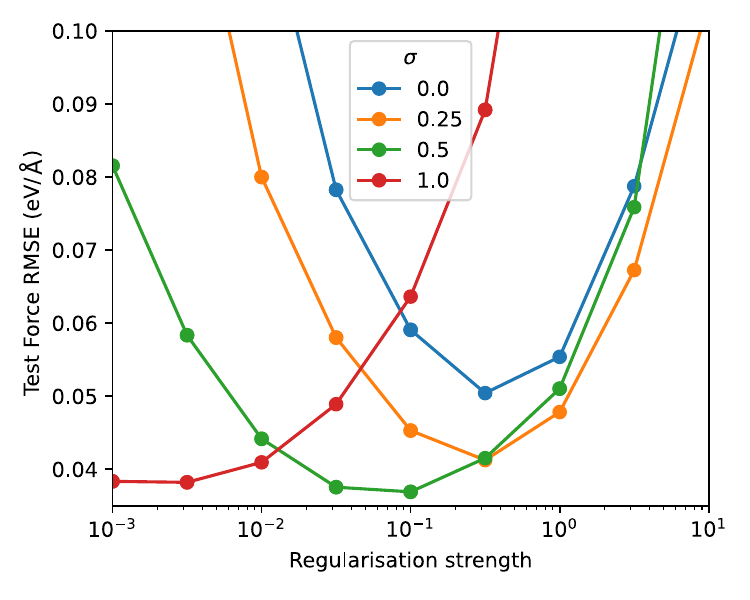}
    \caption{Force RMSE on the Aspirin test set as a function of regularization strength for the ACE potentials fit using $N=4$ and a Gaussian regularity prior with various values for $\sigma$. }
    \label{fig:aspirin_errors}
\end{figure}

The training and test sets used here contained 1000 configurations each and were obtained by drawing samples from the original trajectories at equi-spaced intervals, rather than using fully random splits. ACE potentials were then fit to this training set using smaller basis sets, with polynomial degree=(18,18,12) for the $N=1, 2$ and 3 features respectively, and a larger basis set with $N=4$ and polynomial degree=(18,18,18,12). As before, an Agnesi transformation was applied to the radial coordinate \cite{witt2023acepotentials} and the fits was performed using standard Tikhonov regularization with varying strengths. The best achievable force errors on the test set are tabulated in Table \ref{tab:aspirin_error} and are shown as a function of regularization strength for the $N=4$ potentials in figure \ref{fig:aspirin_errors}. With the smaller basis (N$\leq$3), a very modest 4\% reduction in the force RSME on the test set is observed when using $\sigma=0.5$~\AA{} compared to no regularity prior. However, on increasing the correlation order (and number of basis functions) we see a more significant difference, with $\sigma=0.5$~\AA{} reducing the test error by 27\%. It is also notable that without the regularity prior, the extra basis functions give no reduction in the test error. 

\begin{table}[h]
    \caption{Force RMSE on the Aspirin test set for ACE potentials fit using maximum correlation order $N$=3 and 4 for Gaussian regularity priors with various $\sigma$. }
    \label{tab:aspirin_error}
    \centering
    \begin{tabular}{lll}
    \toprule
     & \multicolumn{2}{l}{ RMSE (meV/\AA)} \\
    $\sigma$ (\AA) & $N$=3                  & $N$=4                 \\
    \toprule
    0.0\quad\quad\quad    & 50.6  \quad\quad\quad                & 50.4                 \\
    0.25  & 48.8                  & 41.2                 \\
    0.5   & 48.0                  & \textbf{36.9 }                \\
    1.0   & 51.9                  & 38.2                \\
    \toprule
    \end{tabular}
\end{table}

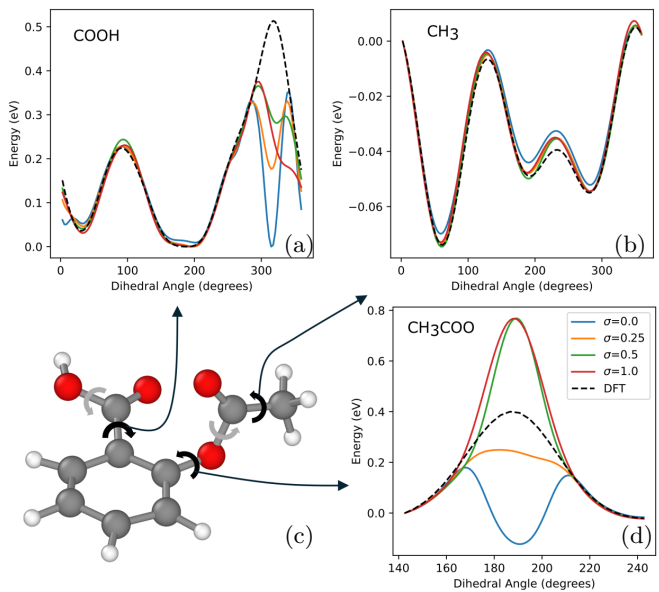
\begin{figure}[htbp]
    \centering
    \begin{tikzpicture}
        \node[anchor=south west, inner sep=0] (img) 
            {\includegraphics[width=0.5 \textwidth]{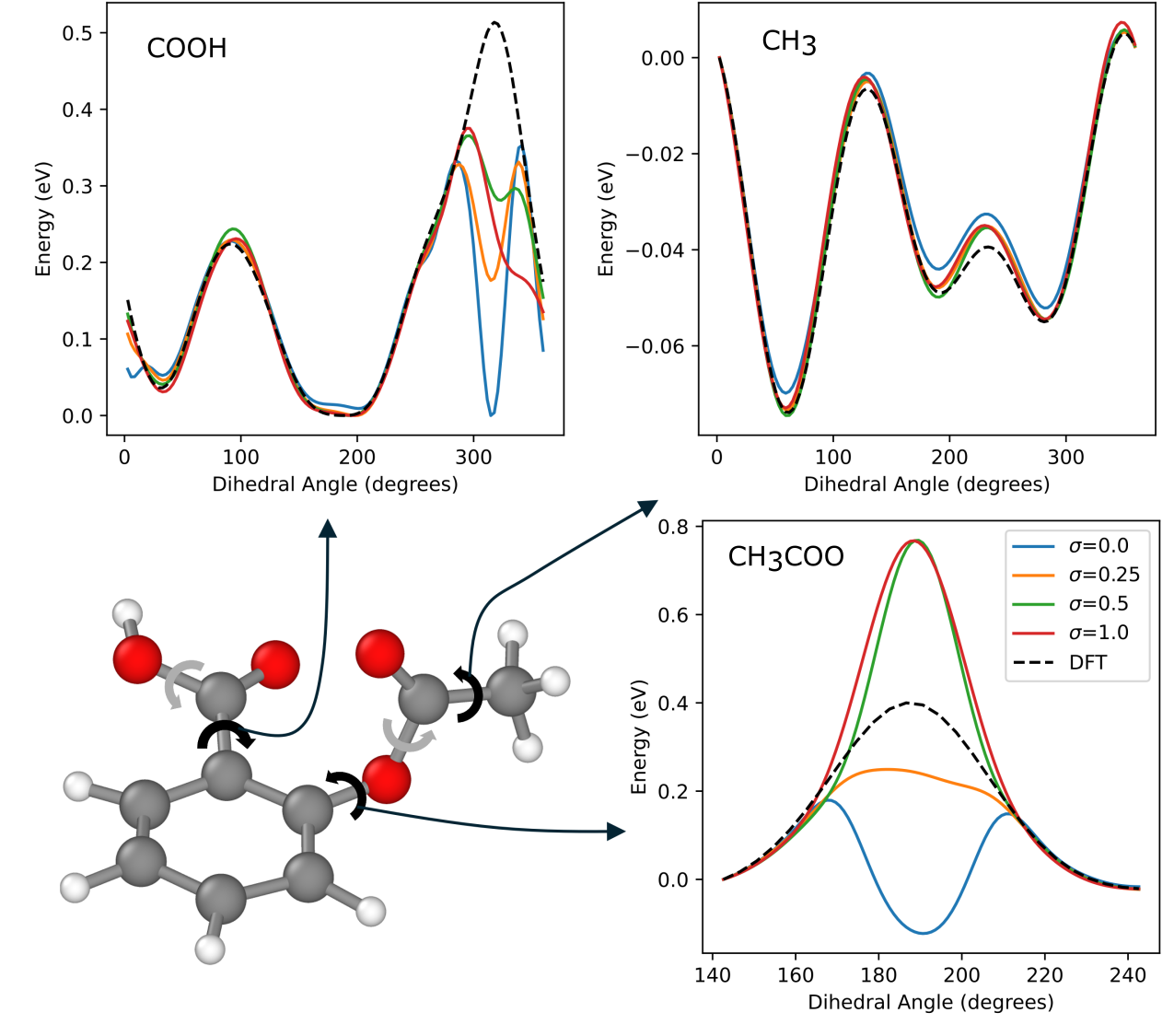}};
        \begin{scope}[x={(img.south east)}, y={(img.north west)}]
            \node[font=\small] at (0.46, 0.6) {(a)};
            \node[font=\small] at (0.95, 0.6) {(b)};
            \node[font=\small] at (0.46, 0.11) {(c)};
            \node[font=\small] at (0.96, 0.11) {(d)};
        \end{scope}
    \end{tikzpicture}
    \caption{(c) A single aspirin molecule is shown with three bonds corresponding to the named functional groups marked in black. Clockwise from the top left the energy of rigidly rotating the COOH (a), \ch{CH3} (b) and \ch{CH3COO} (d) groups are shown as a function of angle. }
    \label{fig:aspirin_dihedral_cuts}
\end{figure}

As with Silicon, we then looked at the effect of the regularity priors on various 1D cuts through the PES. The dimer curves are shown in figure \ref{fig:aspirin_dimers} where it is clear that applying a modest regularity prior helps dampen the spurious high-frequency oscillations seen with $\sigma=0$. Increasing $\sigma$ further down-weights the higher-body order terms relative to the 2-body, and so forces the low polynomial degree 2-body terms to account for ever more of the binding energy. This leads to qualitatively improved behavior for $\sigma \geq 0.25\mathrm{\AA}$, with at most a single minimum seen in each dimer curve. However, due to the lack of dimer data in the training set, the depth of this minimum is not quantitatively correct. Next, we looked at the effect of the regularity prior on higher correlation order terms in the potentials by computing the energy as the COOH, \ch{CH3COO} and \ch{CH3} groups in an aspirin molecule were rigidly rotated. The energy profiles are shown in figure \ref{fig:aspirin_dihedral_cuts}, with the black dashed lines showing the true reference energy computed using ORCA \cite{ORCA} with the same settings as used for the original dataset. All potentials show good agreement with the reference for rotations of the methyl group, which is consistent with the low energy barriers for this rotational mode leading to it being well sampled in the training data. In contrast, there are much larger energy barriers for rotating the other two groups with significant differences between the predicted energies and the reference values. In particular, the $\sigma=0$ potential has a deep minimum where there should be a maximum. Increasing $\sigma$ gradually removes the minimum seen near 320$^\circ$ when rotating the COOH whilst for the \ch{CH3COO} it leads to an overestimation of the energy maximum at 190$^\circ$. 

\begin{figure}[htbp]
    \centering
    \begin{tikzpicture}
        \node[anchor=south west, inner sep=0] (img) 
            {\includegraphics[width=0.5 \textwidth]{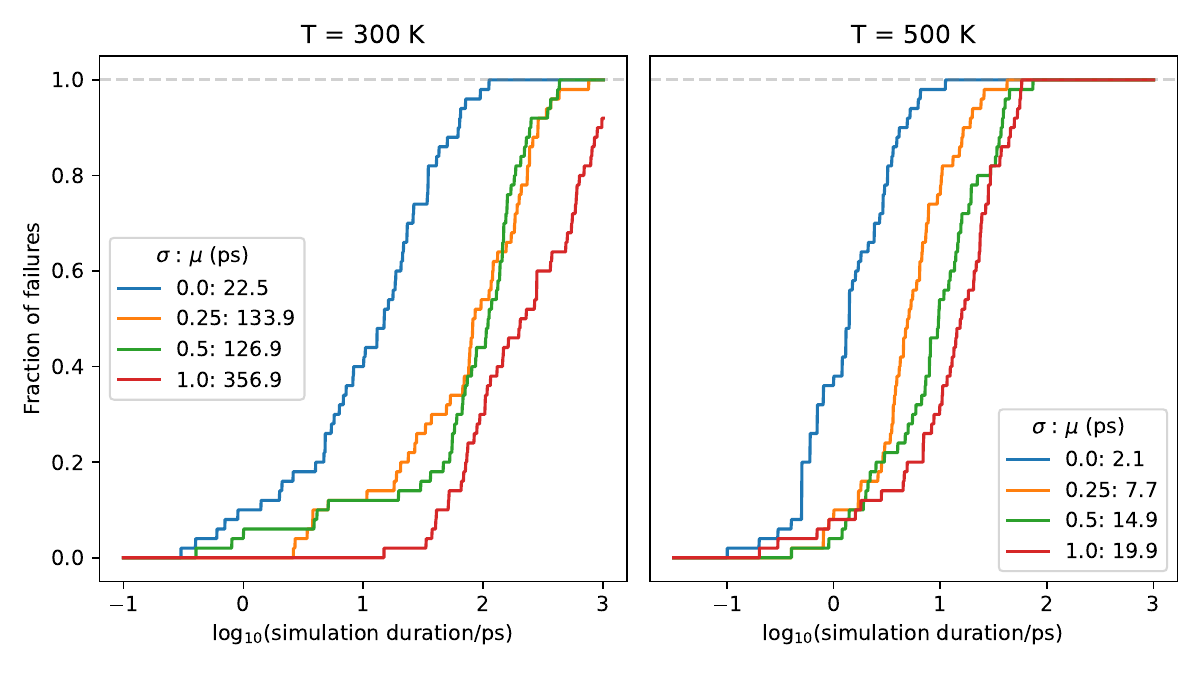}};
        \begin{scope}[x={(img.south east)}, y={(img.north west)}]
            \node[font=\small] at (0.11, 0.85) {(a)};
            \node[font=\small] at (0.57, 0.85) {(b)};
        \end{scope}
    \end{tikzpicture}
    \caption{Duration of MD simulations of a single Aspirin molecule before encountering a ``hole" i.e. catastrophic failure. The simulations were performed at (a) 300 and (b) 500 K and were terminated after 1 ns if failure had not occurred.}
    \label{fig:aspirin_MD_lengths}
\end{figure}

The chronic underestimation of energy by the $\sigma=0$ potentials seen in figure \ref{fig:aspirin_dihedral_cuts} suggests that using larger values of $\sigma$ may lead to more stable \ac{MD} as there appear to be fewer ``holes'' in the PES. To test this we took 100 configurations in the test set and attempted to perform 1 ns of MD at both 300 and 500 K. The dynamics was performed using the \texttt{ase} python package \cite{larsen2017atomic} with a 0.5 fs time step and a Langevin thermostat \cite{vanden2006second} with a friction coefficient of 3 ps$^{-1}$. The simulations were terminated whenever a covalent bond present in the original molecule broke or if a new bond was formed. Thresholds for bonds breaking/forming were determined by analyzing the minimum/maximum distances present between bonded elements in the training set and then adding/subtracting an additional 0.5 \AA~ of tolerance. In practice the details of these criteria were found to be unimportant as any change in the number of bonds was typically accompanied by a dramatic failure of the potential where the molecule exploded. The distribution of the duration of these MD runs is shown in figure \ref{fig:aspirin_MD_lengths} and shows that using the regularity prior led to $~\sim10\times$ longer average simulation times. To verify that this increase was not caused exclusively by over-stabilization of certain configurations the joint distribution of dihedral angles was inspected for trajectories computed with each potential. The distributions are visualized in figure \ref{fig:aspirin_MD_trajectories} and show no evidence that the trajectories with larger $\sigma$ explored less of the configuration space. Furthermore, the minimum energy paths between minima were investigated using NEBs \cite{henkelman2000climbing} and found to be in good agreement with \ac{DFT} for all potentials, see figure \ref{fig:aspirin_neb_paths}. As such, we conclude that applying a regularity prior provides a significant improvement to MD stability. 

\begin{figure}[h!]
    \centering
    \begin{tikzpicture}
        \node[anchor=south west, inner sep=0] (img) 
            {\includegraphics[width=0.5 \textwidth]{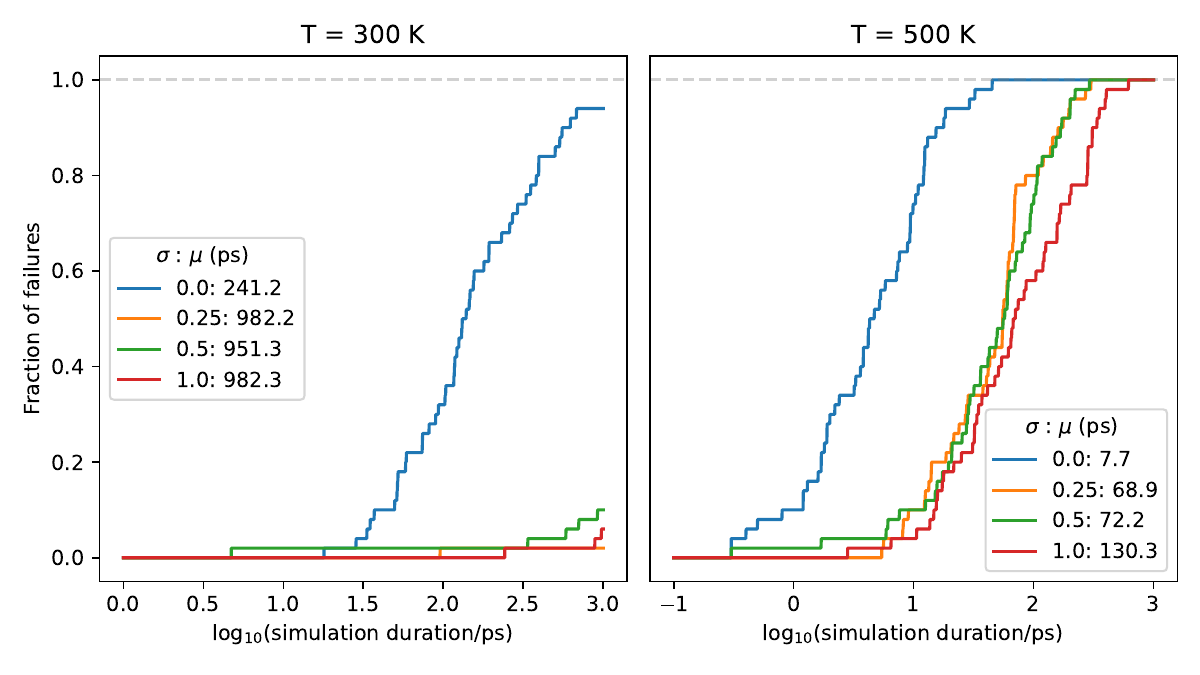}};
        \begin{scope}[x={(img.south east)}, y={(img.north west)}]
            \node[font=\small] at (0.11, 0.85) {(a)};
            \node[font=\small] at (0.57, 0.85) {(b)};
        \end{scope}
    \end{tikzpicture}
    \caption{Duration of MD simulations of a single Aspirin molecule before encountering a ``hole" i.e. catastrophic failure, using the ACE potentials fit to the additional data generated via random dihedral rotations. The simulations were performed at (a) 300 and (b) 500 K and were terminated after 1 ns if failure had not occurred.}
    \label{fig:aspirin_MD_lengths_EE}
\end{figure}

We stress that the aim of this test was not to obtain a robust potential, but instead study the effects of the regularity priors on MD stability. As such, by design we have chosen to use a relatively small amount of training data so that the MD is not stable and the effect of the regularity priors can be easily seen. To demonstrate that the MD stability can easily be improved by adding more data, we generated 10,000 configurations by applying random rotations to all 5 dihedral angles marked in figure \ref{fig:aspirin_dihedral_cuts}. These were then filtered down to $\sim 6500$ configurations using minimum distance constraints before farthest-point-sampling was used to select 284 new configurations to be added to the training set. The entire process was repeated and used to generate an additional 144 configurations which were added to the test set. By generating the data in this way, as opposed to sampling from the MD trajectories generated using the ACE potentials, the new configurations are completely unbiased with respect to any of the potentials, see figure \ref{fig:aspirin_new_distribution} for the energy and force distribution. This enlarged dataset was then refit using the same procedure as described previously. The force errors on the new test set are shown in figure \ref{fig:aspirin_EE_errors} and, as before, applying a regularity prior leads to a significant reduction (40\%) in test force RMSE. Repeating the MD simulations we see a clear improvement in MD stability, with the new durations shown in figure \ref{fig:aspirin_MD_lengths_EE}. Finally, we also note that the additional data leads to greatly improved dihedral  compared to \ac{DFT} seen for all potentials in figure \ref{fig:aspirin_EE_dihedrals}. 

\section{Discussion}
In this work we analyzed the concept of regularity priors for linear ACE models. To summarize in non-technical terms, regularity priors can be understood as follows: ACE basis functions have an associated radial frequency (indexed by $n$) and angular frequency (indexed by $l$), with larger indices corresponding to higher frequencies. Large weights on the high frequency features result in non-smooth potentials. Generalized Tikhonov regularization enforces smaller weights on high frequency terms and thus results in smooth potentials that are significantly more robust in realistic simulation scenarios.

Regularity priors are interpretable from a Bayesian perspective and are implemented as a modified regularization term. 
Equivalently, the priors can be interpreted as a rescaling of the one-particle basis functions. This allows standard regression methods to be applied, and ensures that no additional complexity or computational cost arises during fitting. 

%We tested three different forms of regularity prior based on the predicted potential energy surface having $p$ continuous derivatives, being analytic and an over-regularizing Gaussian prior.  
We tested three different regularity priors: two which either assume that the predicted PES has p continuous derivatives or is analytical, as well as a third with an over-regularizing Gaussian form.
By considering the effect of the Gaussian prior on the fictitious neighbor density it was shown be equivalent to the Gaussian broadening used in SOAP-GAP models. 

Numerical tests indicated that the exact form of the prior made little difference but that using a regularity prior led to significant gains when compared to the standard uniform prior. These tests were performed on the diverse Silicon-GAP-18 dataset, a small subset of the same data, and configurations of individual Aspirin molecules taken from \ac{MD} trajectories.  For all datasets using regularity priors improved the accuracy on the unseen test set, with an almost 30-40\% reduction in force RMSE.  Additional testing also revealed improvements in smoothness along various 1D cuts through the PES, more consistent repulsion at close approach, the removal of many false minima encountered during random structure searching and enhanced stability during \ac{MD} simulations.

We conclude by noting that regularity priors are not a substitute for data, and that using one clearly does not guarantee a potential will behave well in all circumstances. However, they offer consistently improved performance at essentially no additional computational cost or complexity. An interesting avenue for future work is to assess how similar ideas can be incorporated into more complex, non-linear architectures such as a neural-network-based potentials and message-passing architectures such as MACE \cite{batatia2022mace}, NequIP \cite{batzner20223}, PaiNN \cite{schutt2021equivariant} etc. 

\section{Acknowledgments}
ABP acknowledges support from the CASTEP-USER grant funded by UK Research and Innovation (EP/W030438/1). We used computational resources of the UK HPC service ARCHER2 via the UKCP consortium and funded by EPSRC Grant No. EP/P022596/1 and EPSRC Grant No. EP/X035891/1. CO acknowledges the support of the Natural Sciences and Engineering Research Council of Canada (NSERC). J.D.M. acknowledges funding from the EPSRC Centre for Doctoral Training in Inorganic Chemistry for Future Manufacturing (OxICFM), Grant No. EP/S023828/1.

\section{Conflicts of Interest}

GC, CO and JPD are partners in Symmetric Group LLP that licenses force fields commercially. GC also has equity interest in Ångström AI.

\bibliography{references}

@article{bartok2018machine,
  title={Machine learning a general-purpose interatomic potential for silicon},
  author={Bart{\'o}k, Albert P and Kermode, James and Bernstein, Noam and Cs{\'a}nyi, G{\'a}bor},
  journal={Physical Review X},
  volume={8},
  number={4},
  pages={041048},
  year={2018},
  publisher={APS}
}

@article{van2023hyperactive,
  title={Hyperactive learning for data-driven interatomic potentials},
  author={van der Oord, Cas and Sachs, Matthias and Kov{\'a}cs, D{\'a}vid P{\'e}ter and Ortner, Christoph and Cs{\'a}nyi, G{\'a}bor},
  journal={npj Computational Materials},
  volume={9},
  number={1},
  pages={168},
  year={2023},
  publisher={Nature Publishing Group UK London}
}

@article{klawohn2023gaussian,
  title={Gaussian approximation potentials: Theory, software implementation and application examples},
  author={Klawohn, Sascha and Darby, James P and Kermode, James R and Cs{\'a}nyi, G{\'a}bor and Caro, Miguel A and Bart{\'o}k, Albert P},
  journal={The Journal of Chemical Physics},
  volume={159},
  number={17},
  year={2023},
  pages={174108},
  publisher={AIP Publishing}
}

@article{bartok2010gaussian,
  title={Gaussian approximation potentials: The accuracy of quantum mechanics, without the electrons},
  author={Bart{\'o}k, Albert P and Payne, Mike C and Kondor, Risi and Cs{\'a}nyi, G{\'a}bor},
  journal={Physical review letters},
  volume={104},
  number={13},
  pages={136403},
  year={2010},
  publisher={APS}
}

@article{witt2023acepotentials,
  title={ACEpotentials. jl: A Julia implementation of the atomic cluster expansion},
  author={Witt, William C and van der Oord, Cas and Gel{\v{z}}inyt{\.e}, Elena and J{\"a}rvinen, Teemu and Ross, Andres and Darby, James P and Ho, Cheuk Hin and Baldwin, William J and Sachs, Matthias and Kermode, James and others},
  journal={The Journal of Chemical Physics},
  volume={159},
  number={16},
  year={2023},
  publisher={AIP Publishing}
}

@article{morrow2022indirect,
  title={Indirect learning and physically guided validation of interatomic potential models},
  author={Morrow, Joe D and Deringer, Volker L},
  journal={The Journal of Chemical Physics},
  volume={157},
  number={10},
  year={2022},
  publisher={AIP Publishing}
}

@article{deringer2021gaussian,
  title={Gaussian process regression for materials and molecules},
  author={Deringer, Volker L and Bart{\'o}k, Albert P and Bernstein, Noam and Wilkins, David M and Ceriotti, Michele and Cs{\'a}nyi, G{\'a}bor},
  journal={Chemical reviews},
  volume={121},
  number={16},
  pages={10073--10141},
  year={2021},
  publisher={ACS Publications}
}

@article{deringer2021origins,
  title={Origins of structural and electronic transitions in disordered silicon},
  author={Deringer, Volker L and Bernstein, Noam and Cs{\'a}nyi, G{\'a}bor and Ben Mahmoud, Chiheb and Ceriotti, Michele and Wilson, Mark and Drabold, David A and Elliott, Stephen R},
  journal={Nature},
  volume={589},
  number={7840},
  pages={59--64},
  year={2021},
  publisher={Nature Publishing Group UK London}
}

@article{zhang2023atomistic,
  title={Atomistic fracture in bcc iron revealed by active learning of Gaussian approximation potential},
  author={Zhang, Lei and Cs{\'a}nyi, G{\'a}bor and Van Der Giessen, Erik and Maresca, Francesco},
  journal={npj Computational Materials},
  volume={9},
  number={1},
  pages={217},
  year={2023},
  publisher={Nature Publishing Group UK London}
}

@article{erhard2022machine,
  title={A machine-learned interatomic potential for silica and its relation to empirical models},
  author={Erhard, Linus C and Rohrer, Jochen and Albe, Karsten and Deringer, Volker L},
  journal={npj Computational Materials},
  volume={8},
  number={1},
  pages={90},
  year={2022},
  publisher={Nature Publishing Group UK London}
}

@article{marchant2023exploring,
  title={Exploring the configuration space of elemental carbon with empirical and machine learned interatomic potentials},
  author={Marchant, George A and Caro, Miguel A and Karasulu, Bora and P{\'a}rtay, Livia B},
  journal={npj Computational Materials},
  volume={9},
  number={1},
  pages={131},
  year={2023},
  publisher={Nature Publishing Group UK London}
}

@article{mattsson2010first,
  title={First-principles and classical molecular dynamics simulation of shocked polymers},
  author={Mattsson, Thomas R and Lane, J Matthew D and Cochrane, Kyle R and Desjarlais, Michael P and Thompson, Aidan P and Pierce, Flint and Grest, Gary S},
  journal={Physical Review B—Condensed Matter and Materials Physics},
  volume={81},
  number={5},
  pages={054103},
  year={2010},
  publisher={APS}
}

@article{heijmans2020gibbs,
  title={Gibbs ensemble Monte Carlo for reactive force fields to determine the vapor--liquid equilibrium of CO2 and H2O},
  author={Heijmans, Koen and Tranca, Ionut C and Smeulders, David MJ and Vlugt, Thijs JH and Gaastra-Nedea, Silvia V},
  journal={Journal of chemical theory and computation},
  volume={17},
  number={1},
  pages={322--329},
  year={2020},
  publisher={ACS Publications}
}

@article{muser2023interatomic,
  title={Interatomic potentials: Achievements and challenges},
  author={M{\"u}ser, Martin H and Sukhomlinov, Sergey V and Pastewka, Lars},
  journal={Advances in Physics: X},
  volume={8},
  number={1},
  pages={2093129},
  year={2023},
  publisher={Taylor \& Francis}
}

@article{vita2023data,
  title={Data efficiency and extrapolation trends in neural network interatomic potentials},
  author={Vita, Joshua A and Schwalbe-Koda, Daniel},
  journal={Machine Learning: Science and Technology},
  volume={4},
  number={3},
  pages={035031},
  year={2023},
  publisher={IOP Publishing}
}

@article{benoit2020measuring,
  title={Measuring transferability issues in machine-learning force fields: the example of gold--iron interactions with linearized potentials},
  author={Benoit, Magali and Amodeo, Jonathan and Combettes, S{\'e}gol{\`e}ne and Khaled, Ibrahim and Roux, Aur{\'e}lien and Lam, Julien},
  journal={Machine Learning: Science and Technology},
  volume={2},
  number={2},
  pages={025003},
  year={2020},
  publisher={IOP Publishing}
}

@article{sivaraman2020machine,
  title={Machine-learned interatomic potentials by active learning: amorphous and liquid hafnium dioxide},
  author={Sivaraman, Ganesh and Krishnamoorthy, Anand Narayanan and Baur, Matthias and Holm, Christian and Stan, Marius and Cs{\'a}nyi, G{\'a}bor and Benmore, Chris and V{\'a}zquez-Mayagoitia, {\'A}lvaro},
  journal={npj Computational Materials},
  volume={6},
  number={1},
  pages={104},
  year={2020},
  publisher={Nature Publishing Group UK London}
}

@article{fu2025learning,
  title={Learning smooth and expressive interatomic potentials for physical property prediction},
  author={Fu, Xiang and Wood, Brandon M and Barroso-Luque, Luis and Levine, Daniel S and Gao, Meng and Dzamba, Misko and Zitnick, C Lawrence},
  journal={arXiv preprint arXiv:2502.12147},
  year={2025}
}

@article{batatia2023foundation,
  title={A foundation model for atomistic materials chemistry},
  author={Batatia, Ilyes and Benner, Philipp and Chiang, Yuan and Elena, Alin M and Kov{\'a}cs, D{\'a}vid P and Riebesell, Janosh and Advincula, Xavier R and Asta, Mark and Avaylon, Matthew and Baldwin, William J and others},
  journal={arXiv preprint arXiv:2401.00096},
  year={2023}
}

@article{merchant2023scaling,
  title={Scaling deep learning for materials discovery},
  author={Merchant, Amil and Batzner, Simon and Schoenholz, Samuel S and Aykol, Muratahan and Cheon, Gowoon and Cubuk, Ekin Dogus},
  journal={Nature},
  volume={624},
  number={7990},
  pages={80--85},
  year={2023},
  publisher={Nature Publishing Group UK London}
}

@article{rhodes2025orb,
  title={Orb-v3: atomistic simulation at scale},
  author={Rhodes, Benjamin and Vandenhaute, Sander and {\v{S}}imkus, Vaidotas and Gin, James and Godwin, Jonathan and Duignan, Tim and Neumann, Mark},
  journal={arXiv preprint arXiv:2504.06231},
  year={2025}
}

@article{bochkarev2024graph,
  title={Graph atomic cluster expansion for semilocal interactions beyond equivariant message passing},
  author={Bochkarev, Anton and Lysogorskiy, Yury and Drautz, Ralf},
  journal={Physical Review X},
  volume={14},
  number={2},
  pages={021036},
  year={2024},
  publisher={APS}
}

@article{riebesell2025framework,
  title={A framework to evaluate machine learning crystal stability predictions},
  author={Riebesell, Janosh and Goodall, Rhys EA and Benner, Philipp and Chiang, Yuan and Deng, Bowen and Ceder, Gerbrand and Asta, Mark and Lee, Alpha A and Jain, Anubhav and Persson, Kristin A},
  journal={Nature Machine Intelligence},
  volume={7},
  number={6},
  pages={836--847},
  year={2025},
  publisher={Nature Publishing Group UK London}
}

@inproceedings{
    chiang2025mlip,
    title={{MLIP} Arena: Advancing Fairness and Transparency in Machine Learning Interatomic Potentials through an Open and Accessible Benchmark Platform},
    author={Yuan Chiang and Tobias Kreiman and Elizabeth Weaver and Ishan Amin and Matthew Kuner and Christine Zhang and Aaron Kaplan and Daryl Chrzan and Samuel M Blau and Aditi S. Krishnapriyan and Mark Asta},
    booktitle={AI for Accelerated Materials Design - ICLR 2025},
    year={2025},
    url={https://openreview.net/forum?id=ysKfIavYQE}
}

@incollection{ziegler1985stopping,
  title={The stopping and range of ions in matter},
  author={Ziegler, James F and Biersack, Jochen P},
  booktitle={Treatise on heavy-ion science: volume 6: astrophysics, chemistry, and condensed matter},
  pages={93--129},
  year={1985},
  publisher={Springer}
}

@article{nordlund2025repulsive,
  title={Repulsive interatomic potentials calculated at three levels of theory},
  author={Nordlund, Kai and Lehtola, Susi and Hobler, Gerhard},
  journal={Physical Review A},
  volume={111},
  number={3},
  pages={032818},
  year={2025},
  publisher={APS}
}

@article{bartok2013representing,
  title={On representing chemical environments},
  author={Bart{\'o}k, Albert P and Kondor, Risi and Cs{\'a}nyi, G{\'a}bor},
  journal={Physical Review B—Condensed Matter and Materials Physics},
  volume={87},
  number={18},
  pages={184115},
  year={2013},
  publisher={APS}
}

@article{yang2024mattersim,
  title={Mattersim: A deep learning atomistic model across elements, temperatures and pressures},
  author={Yang, Han and Hu, Chenxi and Zhou, Yichi and Liu, Xixian and Shi, Yu and Li, Jielan and Li, Guanzhi and Chen, Zekun and Chen, Shuizhou and Zeni, Claudio and others},
  journal={arXiv preprint arXiv:2405.04967},
  year={2024}
}

@article{neumann2024orb,
  title={Orb: A fast, scalable neural network potential},
  author={Neumann, Mark and Gin, James and Rhodes, Benjamin and Bennett, Steven and Li, Zhiyi and Choubisa, Hitarth and Hussey, Arthur and Godwin, Jonathan},
  journal={arXiv preprint arXiv:2410.22570},
  year={2024}
}

@article{park2024scalable,
  title={Scalable parallel algorithm for graph neural network interatomic potentials in molecular dynamics simulations},
  author={Park, Yutack and Kim, Jaesun and Hwang, Seungwoo and Han, Seungwu},
  journal={Journal of chemical theory and computation},
  volume={20},
  number={11},
  pages={4857--4868},
  year={2024},
  publisher={ACS Publications}
}

@article{barroso2024open,
  title={Open materials 2024 (omat24) inorganic materials dataset and models},
  author={Barroso-Luque, Luis and Shuaibi, Muhammed and Fu, Xiang and Wood, Brandon M and Dzamba, Misko and Gao, Meng and Rizvi, Ammar and Zitnick, C Lawrence and Ulissi, Zachary W},
  journal={arXiv preprint arXiv:2410.12771},
  year={2024}
}

@article{drautz2019atomic,
  title={Atomic cluster expansion for accurate and transferable interatomic potentials},
  author={Drautz, Ralf},
  journal={Physical Review B},
  volume={99},
  number={1},
  pages={014104},
  year={2019},
  publisher={APS}
}

@article{dusson2022atomic,
  title={Atomic cluster expansion: Completeness, efficiency and stability},
  author={Dusson, Genevieve and Bachmayr, Markus and Cs{\'a}nyi, G{\'a}bor and Drautz, Ralf and Etter, Simon and van Der Oord, Cas and Ortner, Christoph},
  journal={Journal of Computational Physics},
  volume={454},
  pages={110946},
  year={2022},
  publisher={Elsevier}
}

@article{bigi2022smooth,
  title={A smooth basis for atomistic machine learning},
  author={Bigi, Filippo and Huguenin-Dumittan, Kevin K and Ceriotti, Michele and Manolopoulos, David E},
  journal={The Journal of Chemical Physics},
  volume={157},
  number={23},
  year={2022},
  publisher={AIP Publishing}
}

@book{evans2022partial,
  title={Partial differential equations},
  author={Evans, Lawrence C},
  volume={19},
  year={2022},
  publisher={American mathematical society}
}

@article{pickard2011ab,
  title={Ab initio random structure searching},
  author={Pickard, Chris J and Needs, RJ},
  journal={Journal of Physics: Condensed Matter},
  volume={23},
  number={5},
  pages={053201},
  year={2011},
  publisher={IOP Publishing}
}

@article{fu2022forces,
  title={Forces are not enough: Benchmark and critical evaluation for machine learning force fields with molecular simulations},
  author={Fu, Xiang and Wu, Zhenghao and Wang, Wujie and Xie, Tian and Keten, Sinan and Gomez-Bombarelli, Rafael and Jaakkola, Tommi},
  journal={arXiv preprint arXiv:2210.07237},
  year={2022}
}

@article{christensen2020role,
  title={On the role of gradients for machine learning of molecular energies and forces},
  author={Christensen, Anders S and Von Lilienfeld, O Anatole},
  journal={Machine Learning: Science and Technology},
  volume={1},
  number={4},
  pages={045018},
  year={2020},
  publisher={IOP Publishing}
}

@article{ORCA,
author = {Neese,F.},
title = {The ORCA program system},
journal = {Wiley Interdiscip. Rev. Comput. Mol. Sci.},
volume = {2},
number = {1},
pages = {73-78},
DOI = {10.1002/wcms.81},
year = {2012},
type = {journal Article}
}

@article{henkelman2000climbing,
  title={A climbing image nudged elastic band method for finding saddle points and minimum energy paths},
  author={Henkelman, Graeme and Uberuaga, Blas P and J{\'o}nsson, Hannes},
  journal={The Journal of chemical physics},
  volume={113},
  number={22},
  pages={9901--9904},
  year={2000},
  publisher={American Institute of Physics}
}

@article{kaufmann1989single,
  title={Single-centre expansion of Gaussian basis functions and the angular decomposition of their overlap integrals},
  author={Kaufmann, KetWBAUMEISTER and Baumeister, W},
  journal={Journal of Physics B: Atomic, Molecular and Optical Physics},
  volume={22},
  number={1},
  pages={1},
  year={1989},
  publisher={IOP Publishing}
}

@book{watson1922treatise,
  title={A treatise on the theory of Bessel functions},
  author={Watson, George Neville},
  volume={3},
  year={1922},
  publisher={The University Press}
}

@article{kresse1996efficiency,
  title={Efficiency of ab-initio total energy calculations for metals and semiconductors using a plane-wave basis set},
  author={Kresse, Georg and Furthm{\"u}ller, J{\"u}rgen},
  journal={Computational materials science},
  volume={6},
  number={1},
  pages={15--50},
  year={1996},
  publisher={Elsevier}
}

@article{vanden2006second,
  title={Second-order integrators for Langevin equations with holonomic constraints},
  author={Vanden-Eijnden, Eric and Ciccotti, Giovanni},
  journal={Chemical physics letters},
  volume={429},
  number={1-3},
  pages={310--316},
  year={2006},
  publisher={Elsevier}
}

@article{larsen2017atomic,
  title={The atomic simulation environment—a Python library for working with atoms},
  author={Larsen, Ask Hjorth and Mortensen, Jens J{\o}rgen and Blomqvist, Jakob and Castelli, Ivano E and Christensen, Rune and Du{\l}ak, Marcin and Friis, Jesper and Groves, Michael N and Hammer, Bj{\o}rk and Hargus, Cory and others},
  journal={Journal of Physics: Condensed Matter},
  volume={29},
  number={27},
  pages={273002},
  year={2017},
  publisher={IOP Publishing}
}

@article{hohenberg1964inhomogeneous,
  title={Inhomogeneous electron gas},
  author={Hohenberg, Pierre and Kohn, Walter},
  journal={Physical review},
  volume={136},
  number={3B},
  pages={B864},
  year={1964},
  publisher={APS}
}

@article{kohn1965self,
  title={Self-consistent equations including exchange and correlation effects},
  author={Kohn, Walter and Sham, Lu Jeu},
  journal={Physical review},
  volume={140},
  number={4A},
  pages={A1133},
  year={1965},
  publisher={APS}
}

@article{behler2007generalized,
  title={Generalized neural-network representation of high-dimensional potential-energy surfaces},
  author={Behler, J{\"o}rg and Parrinello, Michele},
  journal={Physical review letters},
  volume={98},
  number={14},
  pages={146401},
  year={2007},
  publisher={APS}
}

@article{nandi2019using,
  title={Using gradients in permutationally invariant polynomial potential fitting: A demonstration for CH4 using as few as 100 configurations},
  author={Nandi, Apurba and Qu, Chen and Bowman, Joel M},
  journal={Journal of Chemical Theory and Computation},
  volume={15},
  number={5},
  pages={2826--2835},
  year={2019},
  publisher={ACS Publications}
}

@article{van2020regularised,
  title={Regularised atomic body-ordered permutation-invariant polynomials for the construction of interatomic potentials},
  author={van Der Oord, Cas and Dusson, Genevieve and Cs{\'a}nyi, G{\'a}bor and Ortner, Christoph},
  journal={Machine Learning: Science and Technology},
  volume={1},
  number={1},
  pages={015004},
  year={2020},
  publisher={IOP Publishing}
}

@article{deringer2018data,
  title={Data-driven learning of total and local energies in elemental boron},
  author={Deringer, Volker L and Pickard, Chris J and Cs{\'a}nyi, G{\'a}bor},
  journal={Physical review letters},
  volume={120},
  number={15},
  pages={156001},
  year={2018},
  publisher={APS}
}

@article{batatia2022mace,
  title={MACE: Higher order equivariant message passing neural networks for fast and accurate force fields},
  author={Batatia, Ilyes and Kovacs, David P and Simm, Gregor and Ortner, Christoph and Cs{\'a}nyi, G{\'a}bor},
  journal={Advances in neural information processing systems},
  volume={35},
  pages={11423--11436},
  year={2022}
}

@article{batzner20223,
  title={E (3)-equivariant graph neural networks for data-efficient and accurate interatomic potentials},
  author={Batzner, Simon and Musaelian, Albert and Sun, Lixin and Geiger, Mario and Mailoa, Jonathan P and Kornbluth, Mordechai and Molinari, Nicola and Smidt, Tess E and Kozinsky, Boris},
  journal={Nature communications},
  volume={13},
  number={1},
  pages={2453},
  year={2022},
  publisher={Nature Publishing Group UK London}
}

@inproceedings{schutt2021equivariant,
  title={Equivariant message passing for the prediction of tensorial properties and molecular spectra},
  author={Sch{\"u}tt, Kristof and Unke, Oliver and Gastegger, Michael},
  booktitle={International conference on machine learning},
  pages={9377--9388},
  year={2021},
  organization={PMLR}
}

@article{methfessel1989high,
  title={High-precision sampling for Brillouin-zone integration in metals},
  author={Methfessel, MPAT and Paxton, Anthony T},
  journal={physical review B},
  volume={40},
  number={6},
  pages={3616},
  year={1989},
  publisher={APS}
}

@article{clark2005first,
  title={First principles methods using CASTEP},
  author={Clark, Stewart J and Segall, Matthew D and Pickard, Chris J and Hasnip, Phil J and Probert, Matt IJ and Refson, Keith and Payne, Mike C},
  journal={Zeitschrift f{\"u}r kristallographie-crystalline materials},
  volume={220},
  number={5-6},
  pages={567--570},
  year={2005},
  publisher={De Gruyter Oldenbourg}
}

@article{liu1989limited,
  title={On the limited memory BFGS method for large scale optimization},
  author={Liu, Dong C and Nocedal, Jorge},
  journal={Mathematical programming},
  volume={45},
  number={1},
  pages={503--528},
  year={1989},
  publisher={Springer}
}

@article{zhou2025full,
  title={Full-cycle device-scale simulations of memory materials with a tailored atomic-cluster-expansion potential},
  author={Zhou, Yuxing and Thomas du Toit, Daniel F and Elliott, Stephen R and Zhang, Wei and Deringer, Volker L},
  journal={Nature Communications},
  volume={16},
  number={1},
  pages={1--12},
  year={2025},
  publisher={Nature Publishing Group}
}

@article{deb2001pressure,
  title={Pressure-induced amorphization and an amorphous--amorphous transition in densified porous silicon},
  author={Deb, Sudip K and Wilding, Martin and Somayazulu, Maddury and McMillan, Paul F},
  journal={Nature},
  volume={414},
  number={6863},
  pages={528--530},
  year={2001},
  publisher={Nature Publishing Group UK London}
}

@article{mcmillan2005density,
  title={A density-driven phase transition between semiconducting and metallic polyamorphs of silicon},
  author={McMillan, Paul F and Wilson, Mark and Daisenberger, Dominik and Machon, Denis},
  journal={Nature materials},
  volume={4},
  number={9},
  pages={680--684},
  year={2005},
  publisher={Nature Publishing Group UK London}
}

@article{morrow2023validate,
  title={How to validate machine-learned interatomic potentials},
  author={Morrow, Joe D and Gardner, John LA and Deringer, Volker L},
  journal={The Journal of chemical physics},
  volume={158},
  number={12},
  year={2023},
  pages={121501},
  publisher={AIP Publishing}
}

@article{pandey2011pressure,
  title={Pressure induced crystallization in amorphous silicon},
  author={Pandey, KK and Garg, Nandini and Shanavas, KV and Sharma, Surinder M and Sikka, SK},
  journal={Journal of Applied Physics},
  volume={109},
  number={11},
  year={2011},
  pages={113511},
  publisher={AIP Publishing}
}

@article{ho2024atomic,
  title={Atomic cluster expansion without self-interaction},
  author={Ho, Cheuk Hin and Gutleb, Timon S and Ortner, Christoph},
  journal={Journal of Computational Physics},
  volume={515},
  pages={113271},
  year={2024},
  publisher={Elsevier}
}

\clearpage

\section{Appendix}

\subsection{Priors for the Symmetrized Basis}
\label{sec:symbasis_priors}
In our simplified ACE formulation we wrote 
\begin{equation}
    E_i = \sum_{\nn\ll\mm} c_{\nn\ll\mm} \prod_t A_{i, n_t l_t m_t},
\end{equation}
with summation over all $\nn\ll\mm$ tuples, not only ordered ones. This simplified the presentation of regularity priors. In practice, ACE is implemented by first noticing that only summation over ordered tuples is required, 
\begin{equation}
    \begin{aligned}
    E_i 
    &= 
    \sum_{\nn\ll\mm \text{ ordered}} 
    k_{\nn\ll\mm} c_{\nn\ll\mm} \prod_t A_{i, n_t l_t m_t},  
    \\ 
    &= 
    \sum_{\nn\ll\mm \text{ ordered}}
    k_{\nn\ll\mm} c_{\nn\ll\mm}
    \bAA_{i,\nn\ll\mm} \\ 
    &= 
    ({\bm k} \odot \cc) \cdot \bAA_i,
    \end{aligned}
\end{equation}
where $k_{\nn\ll\mm}$ denotes the number of repeated basis functions (a multi-combination and explicitly computable). Next, one symmetrizes the basis ${\bm B}_i = \mathcal{C} \bAA_i$. 
Because $\cc$, and hence ${\bm k} \odot \cc$, result in an invariant model it follows that ${\bm k} \odot \cc = \mathcal{C}^T {\bm \theta}$ for some parameter vector ${\bm \theta}$ and hence 
\begin{equation}
    E_i = ({\bm k} \odot \cc) \cdot \bAA_i = {\bm \theta} \cdot {\bm B}_i.
\end{equation}
We call ${\bm B}_i$ the {\em ACE basis} or the {\em ACE features}. 

Under the same transformation, the Tikhonov regularizer becomes, 
\begin{equation}
    \| \Gamma \cc \|^2
    = 
    \big\| \Gamma_{\rm ord} ({\bm k} \odot \cc) \big\|^2 
    = \big\| \Gamma_{\rm ord} \mathcal{C}^T {\bm \theta} \big\|^2, 
\end{equation}
where the diagonal entries of $\Gamma_{\rm ord}$ are simply those of $\Gamma$ but restricted to ordered tuples. 
Equivalently, the prior $\cc \sim \mathcal{N}({\bm 0}, \Gamma^{-2})$ transforms to a prior on ${\bm \theta}$, 
\begin{equation}
    {\bm \theta} \sim \mathcal{N}\big({\bm 0}, (\mathcal{C} \Gamma^2 \mathcal{C}^T)^{-1} \big).
\end{equation}
Because of the block-structure of $\mathcal{C}$ one can in fact rewrite $\Gamma_{\rm ord} \mathcal{C}^T {\bm \theta} = \bar{\Gamma} {\bm \theta}$, where $\bar\Gamma$ is diagonal.

\subsection{SOAP-GAP connection}

\label{sec:SOAP_explicit}
Both ACE and SOAP features can be understood as symmetrized tensor-products ($N$ products for $N+1$ body-order features) of a neighbour density. 
In SOAP each neighbour atom is represented using a Gaussian of width $\sigma$, 
\begin{equation}
    c_{i,znlm} = \sum_{j \in \mathcal{N}(i)} 
        \int g_\alpha(\rr - \rr_{ij}) 
            \phi_{znlm}(\rr)
        \, d\rr,
\end{equation}
where $g_\alpha(\rr) = \exp(- \alpha |\rr|^2)$, 
whilst the ACE potentials effectively use a delta function.  
\begin{equation}
    A_{i,znlm} 
    = \sum_{j \in \mathcal{N}(i)}  
        \int \delta(\rr - \rr_{ij}) 
            \phi_{znlm}(\rr) 
        \,d\rr. 
\end{equation}
It turns out, as we show next, that under the specific choice of the one-particle basis proposed in \cite{bigi2022smooth}, applying the Gaussian prior~\eqref{eq:gauss_prior} to the ACE features transforms them into SOAP features. 
Specifically, the self-interacting \ac{ACE} features computed using the \ac{LE} of ref. \cite{bigi2022smooth} as the radial basis are equivalent to SOAP features computed without the central atom contributing to the neighbor density (i.e. \texttt{central\_weight=0} in the GAP code \cite{klawohn2023gaussian}). 

To show this we start by noting that, due to the sifting property of $\delta(\rr-\rr_j)$, the $A_{i, znlm}$ in self-interacting ACE can viewed as the density expansion coefficients of the delta function neighbour density,
\begin{align}
        A_{i,znlm} &= \sum_{j \in \mathcal{N}(i)} \int \diff{\rr} \delta(\rr_j - \rr) R_{nl}(r) Y_l^m(\hat{\rr}) \delta_{zZ_j}\\
         &=\sum_{j \in \mathcal{N}(i)} \phi_{nlm}(\rr_j) = \sum_{j \in \mathcal{N}(i)} R_{nl}(r) Y_l^m(\hat{\rr}) \delta_{zZ_j}
\end{align}
Next we compute the density expansion coefficient $c_{i,nlm}$ for a single neighbour atom represented as a normalised Gaussian of width $\sigma$ centered at $\rr_i$ using the \ac{LE}. 
\begin{equation}
        c_{i,nlm} = \int   \phi_{nlm} \exp(-\alpha |\mathbf{r}-\mathbf{r}_i|^2) r^2 dr d\hat{r} \\
\end{equation}
The \ac{LE} are defined as 
\begin{equation}
    R_{nl}(r) = N_{nl} j_l(z_{nl}a^{-1} r)
\end{equation}
where $j_l(x)$ is a spherical Bessel function of the first kind, $z_{nl}$ is the $n$th zero of $j_l(x)$, $a=r_\mathrm{cut}$ is the cutoff radius and 
\begin{equation}
N_{nl} = \sqrt{\frac{a^3}{2}} j_{l+1}(z_{nl}) 
\end{equation}
is a normalization constant such that 
\begin{equation}
    \int_0^a \diff{r} r^2 R_{nl}(r) R_{n'l}(r) = \delta_{nn'}.
\end{equation}

Using the same approach as in the original SOAP paper \cite{bartok2013representing} we can then evaluate $c_{i,nlm}$ as 

\begin{align*} 
    c_{i,nlm} & = \int   \phi_{nlm} \exp(-\alpha |\mathbf{r}-\mathbf{r}_i|^2) r^2 dr d\hat{r} \\
    & =  \int N_{nl}  j_l(z_{nl}a^{-1}r) Y_{lm}(\hat{r}) \cdot 4 \pi \exp(-\alpha (r^2+r_i^2)) \\
    &  \sum_{l' m'} l_{l'}(2\alpha r r_i) Y_{l'm'}(\hat{r}) Y^*_{lm'}(\hat{r_i}) r^2 dr d\hat{r} \\
    &= 4 \pi N_{nl} \exp(-\alpha r_i^2) Y_{l'm'}(\hat{r_i}) \sum_{l' m'} \int Y_{lm}(\hat{r}) Y^*_{l'm'}(\hat{r})d\hat{r} \\ 
    & \int j_l(z_{nl}a^{-1}r) \exp(-\alpha r^2) l_{l'}(2\alpha r r_i) r^2 dr
\end{align*}

where $l_l(r)$ is a modified spherical Bessel function of the first kind that satisfies $l_l(x) = i^{-l} j_l(ix)$.
Using the normalization of the spherical harmonics,  $ \int Y_{lm}(\hat{r}) Y^*_{l'm'}(\hat{r})d\hat{r} = \delta_{ll'}\delta_{mm'}$ this simplifies to
\begin{align}
    c^i_{nlm} & = 4 \pi N_{nl} \exp(-\alpha r_i^2) Y_{lm}(\hat{r}_i) \cdot   I
\end{align}

where 

\begin{equation}
    I = \int_0^\infty j_l(z_{nl}a^{-1}r) \exp(-\alpha r^2) l_{l}(2\alpha r r_i) r^2 dr
\end{equation}

Note that the upper integration limit is $\infty$, not $r_\mathrm{cut}=a$, as the cutoff function handles atoms entering and leaving the environment. This is a modified version of Weber's second exponential integral, see 13.31 of ref. \cite{watson1922treatise}, which is used in Equation 12 of ref. \cite{kaufmann1989single} with complex $u$ and $v$. Using these results we evaluate $I$ as 

\begin{equation}
    I = \frac{1}{4}\sqrt{\frac{\pi}{\alpha^3}} j_l\left(\frac{z_{nl}r_i}{a}\right) \cdot \exp\left(-\frac{z_{nl}^2}{4\alpha a^2}\right) \cdot \exp \left( \alpha r_i^2\right)
\end{equation}

\begin{figure}[htbp]
    \centering
    \includegraphics[width=0.5\textwidth]{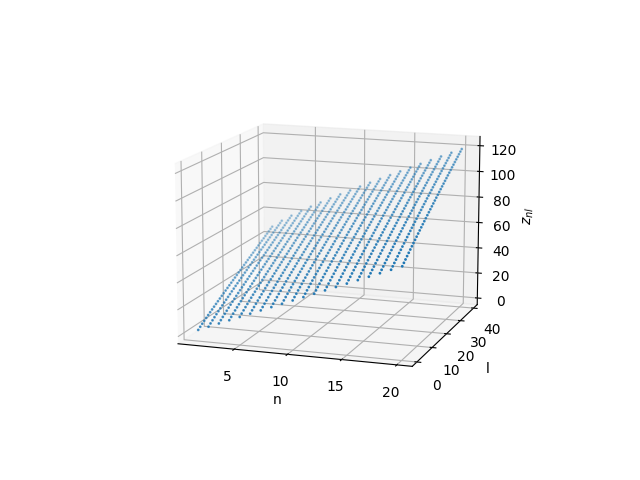}
    \caption{$z_{nl}$ as function of $n$ and $l$}
    \label{fig:znl}
\end{figure}

Combining everything together we arrive at 

\begin{align}
     c^i_{nlm} &=  N_{nl} \cdot j_l\left(\frac{z_{nl}}{a} r_i\right) Y_{lm}(\hat{r}_i) \cdot \exp\left(-\frac{\sigma^2}{2} E_{nl}\right) \\
     &= \phi_{nlm}(\rr_i) \cdot \exp\left(-\frac{\sigma^2}{2} E_{nl}\right) 
\end{align}

where $E_{nl} = z_{nl}^2/a^2$. There is no analytic expression for $z_{nl}$ but it can be crudely approximated as $z_{nl} \approx a n + b l$ as shown in Fig. \ref{fig:znl}.

Under this approximation we see that broadening of the ACE features can be accomplished with a Gaussian prior of the form,

\begin{equation}
    \gamma(\nn \ll) = \prod_i \exp\left(\frac{\sigma^2}{a^2} \cdot  \left[5n_i^2 + 4n_il_i + l_i^2\right]\right)
    \label{eq:exp_reg}
\end{equation}

where, given that we care about the scaling, the coefficients have been rounded to integer values for simplicity.

\clearpage 

\onecolumngrid
\section*{Supplementary}

\setcounter{figure}{0}
\setcounter{table}{0}
\setcounter{page}{1}

\renewcommand{\theequation}{S\arabic{equation}}
\renewcommand{\thefigure}{S\arabic{figure}}
\renewcommand{\thetable}{S\arabic{table}}
\renewcommand{\theHfigure}{Supplement.\thefigure}

\subsection{Si10pc}

\begin{figure*}[h!]
    \centering
    \includegraphics[width=1.0\textwidth]{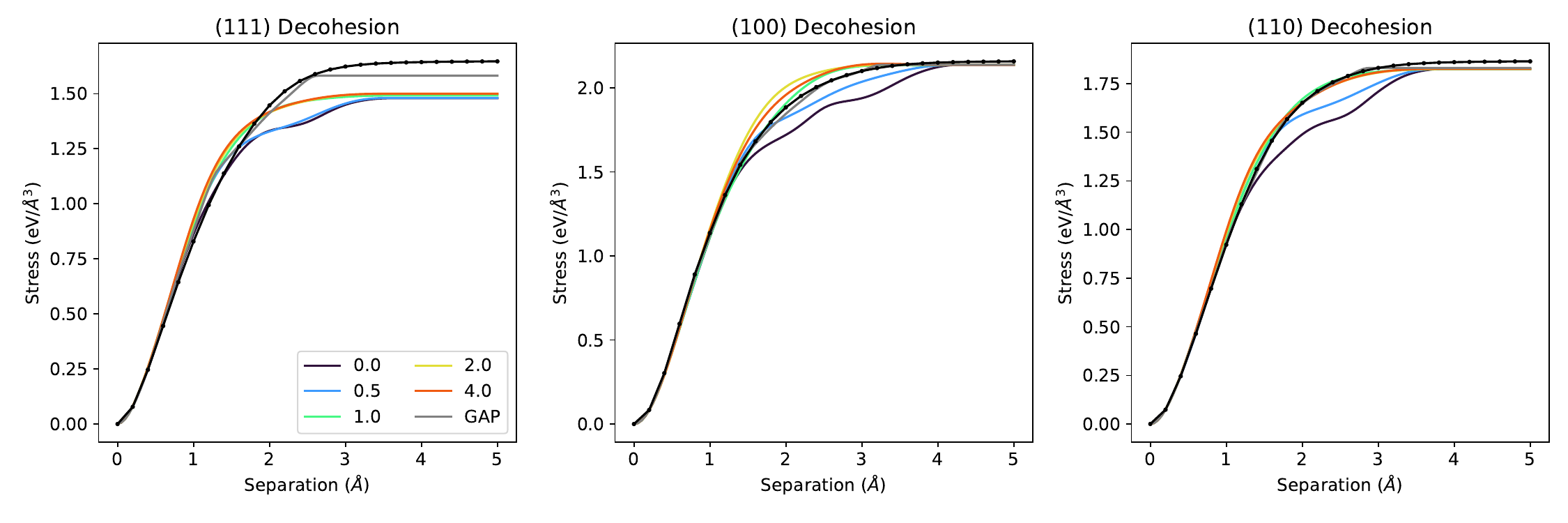}
    \caption{Surface energies as a function of distance for rigid decohesion of the (100), (110) and (111) surfaces of Silicon diamond. The \ac{DFT} reference is shown in black.}
    \label{fig:10_90_gaussian_decohesion_E}
\end{figure*}

\begin{figure}
    \centering
    \includegraphics[width=0.6\textwidth]{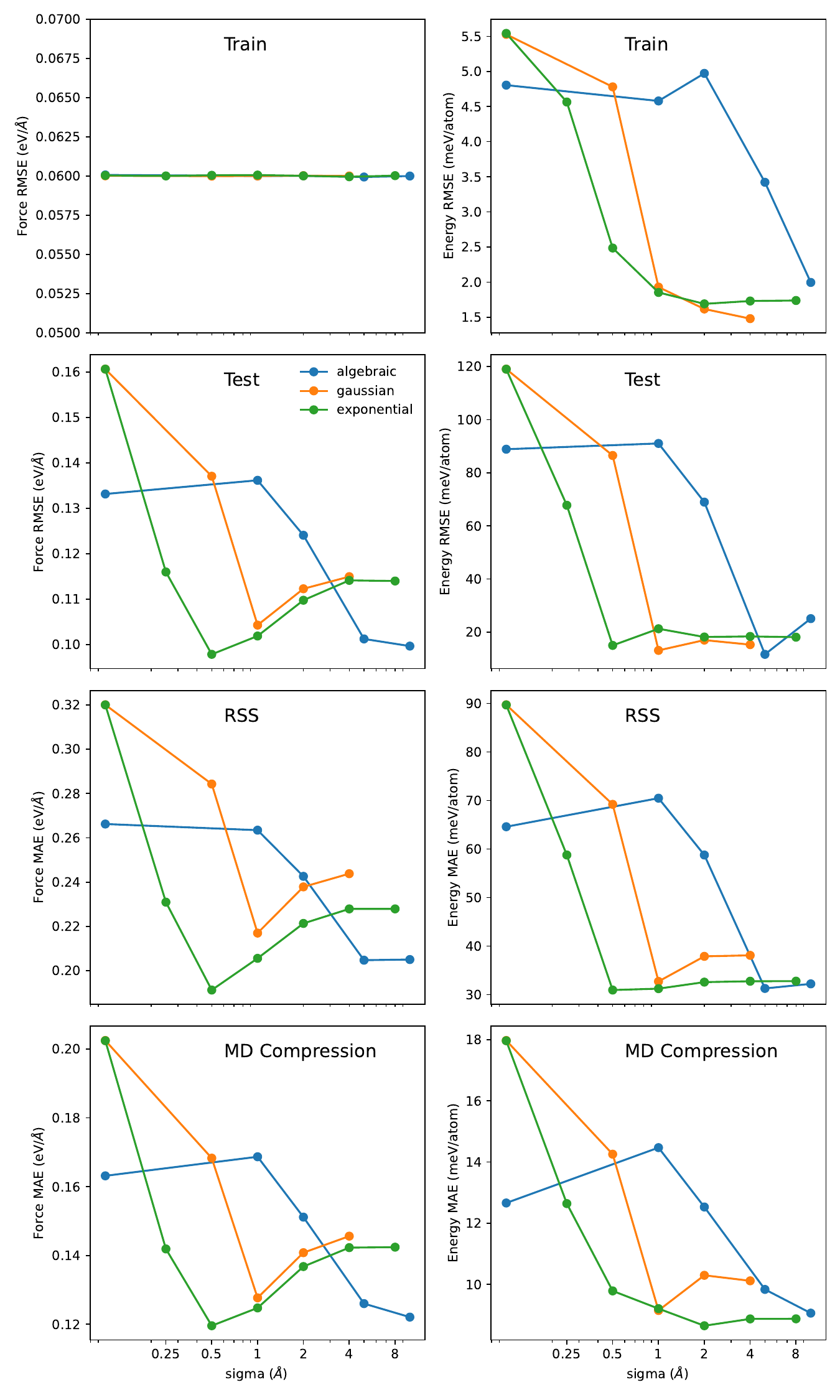}
    \caption{Energy and force errors on four different configuration sets are shown for potentials fit using algebraic, exponential and Gaussian regularity priors. The strength of the regularisation was chosen by tuning the force RMSE on the training set to be 0.06 eV/\AA.}
    \label{fig:10_90_prior_comp}
\end{figure}

\begin{figure*}
    \centering
    \includegraphics[width=\textwidth]{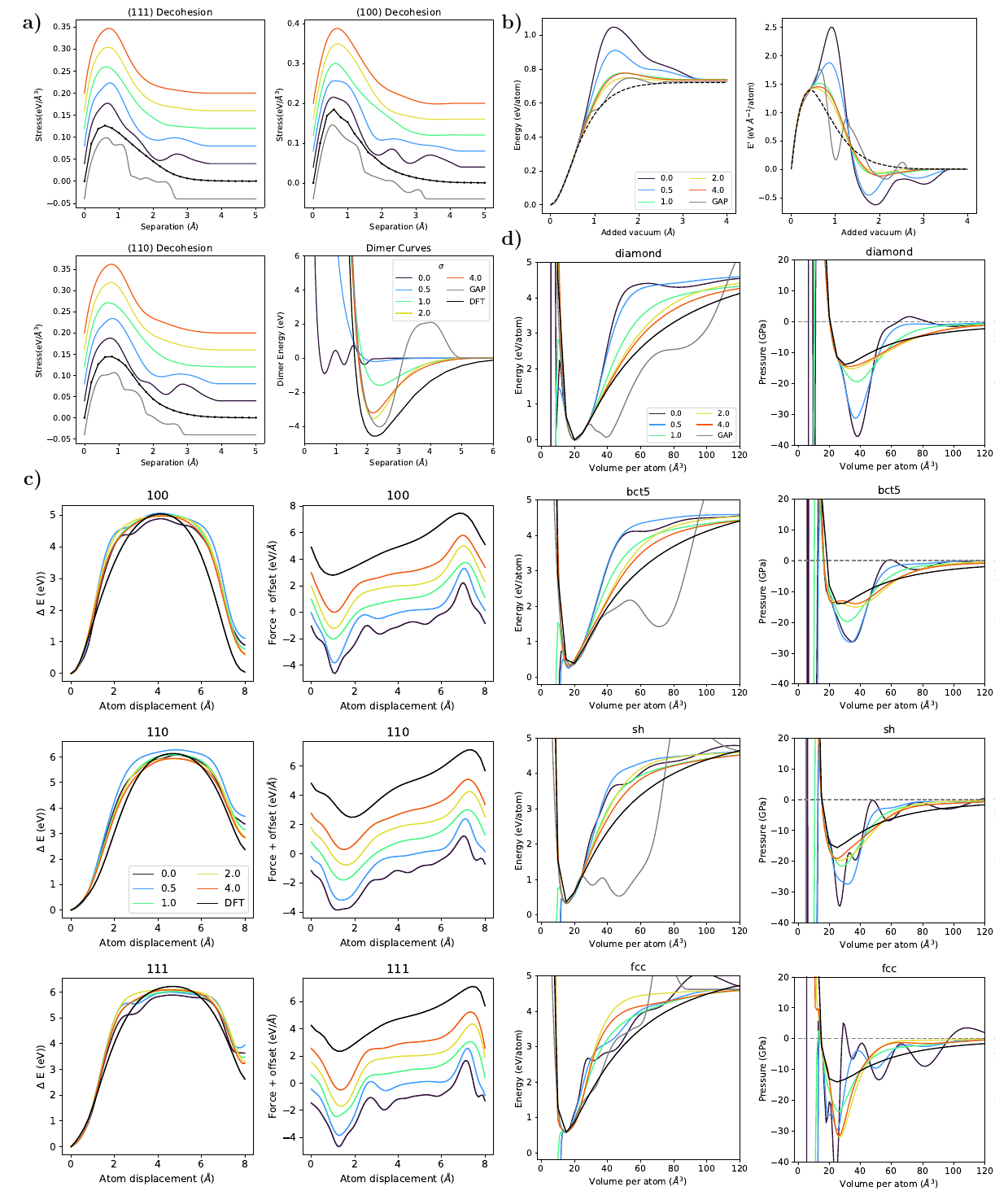}
    \caption{A summary of the tests described in the main text is shown for potentials fit to \textbf{Si10pc} using a \textbf{Gaussian} regularity prior. Specifically the diagrams show: a) the stress during the decohesion of Silicon diamond to various surfaces as well as the corresponding dimer curves. b) the effect of exfoliating of silicon diamond into silicene layers c) what happen when an atom is move across the vacuum from the lower surface of a silicon diamond slab to the upper surface d) the energy (left), and pressure (right) vs volume as various silicon polymorphs are expanded isotropically.}
    \label{fig:10_90_gaussian}
\end{figure*}

\begin{figure*}
    \centering
    \includegraphics[width=\textwidth]{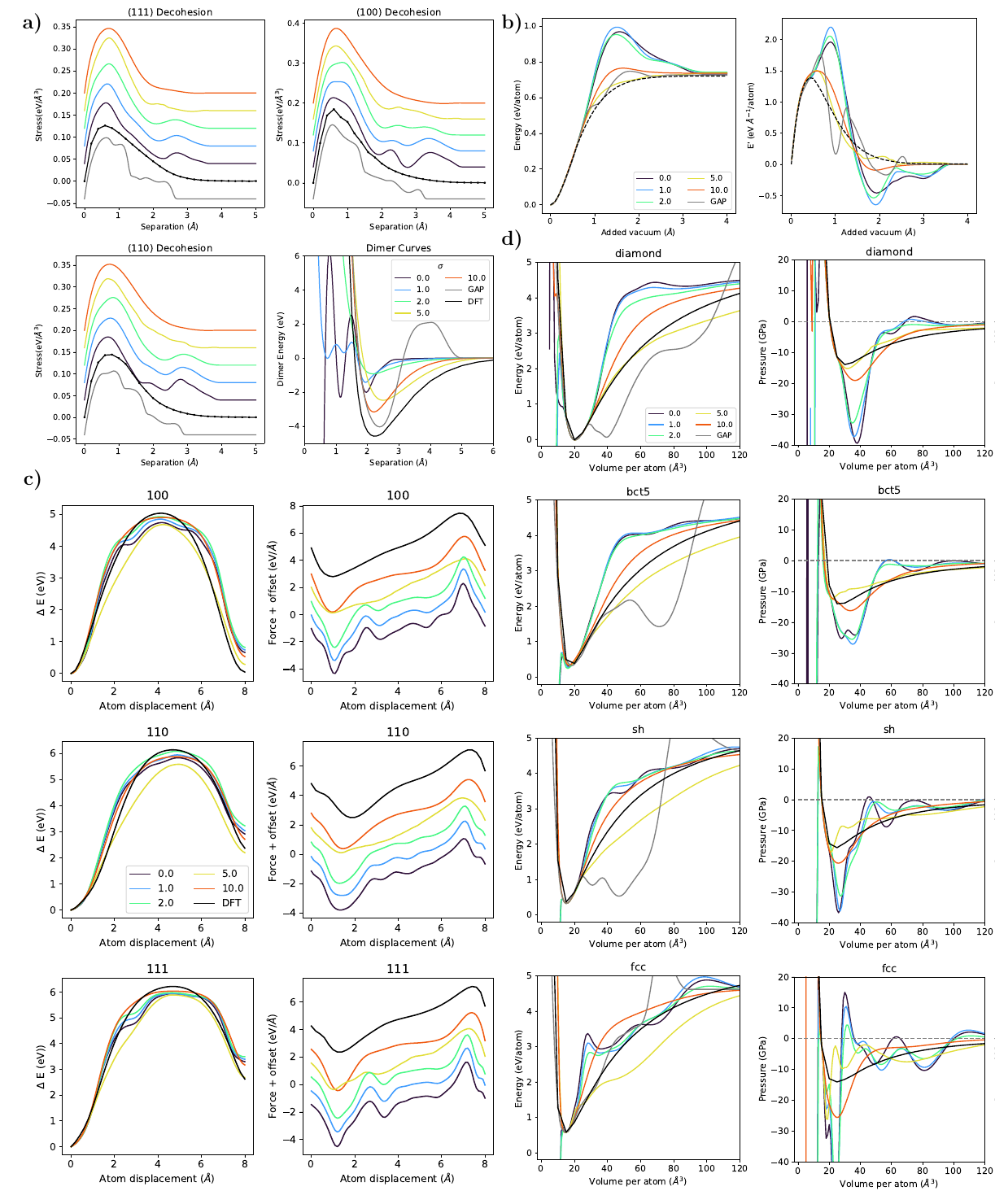}
    \caption{A summary of the tests described in the main text is shown for potentials fit to \textbf{Si10pc} using a \textbf{Gaussian} regularity prior. Specifically the diagrams show: a) the stress during the decohesion of Silicon diamond to various surfaces as well as the corresponding dimer curves. b) the effect of exfoliating of silicon diamond into silicene layers c) what happen when an atom is move across the vacuum from the lower surface of a silicon diamond slab to the upper surface d) the energy (left), and pressure (right) vs volume as various silicon polymorphs are expanded isotropically.}
    \label{fig:10_90_algebraic}
\end{figure*}

\begin{figure*}
    \centering
    \includegraphics[width=\textwidth]{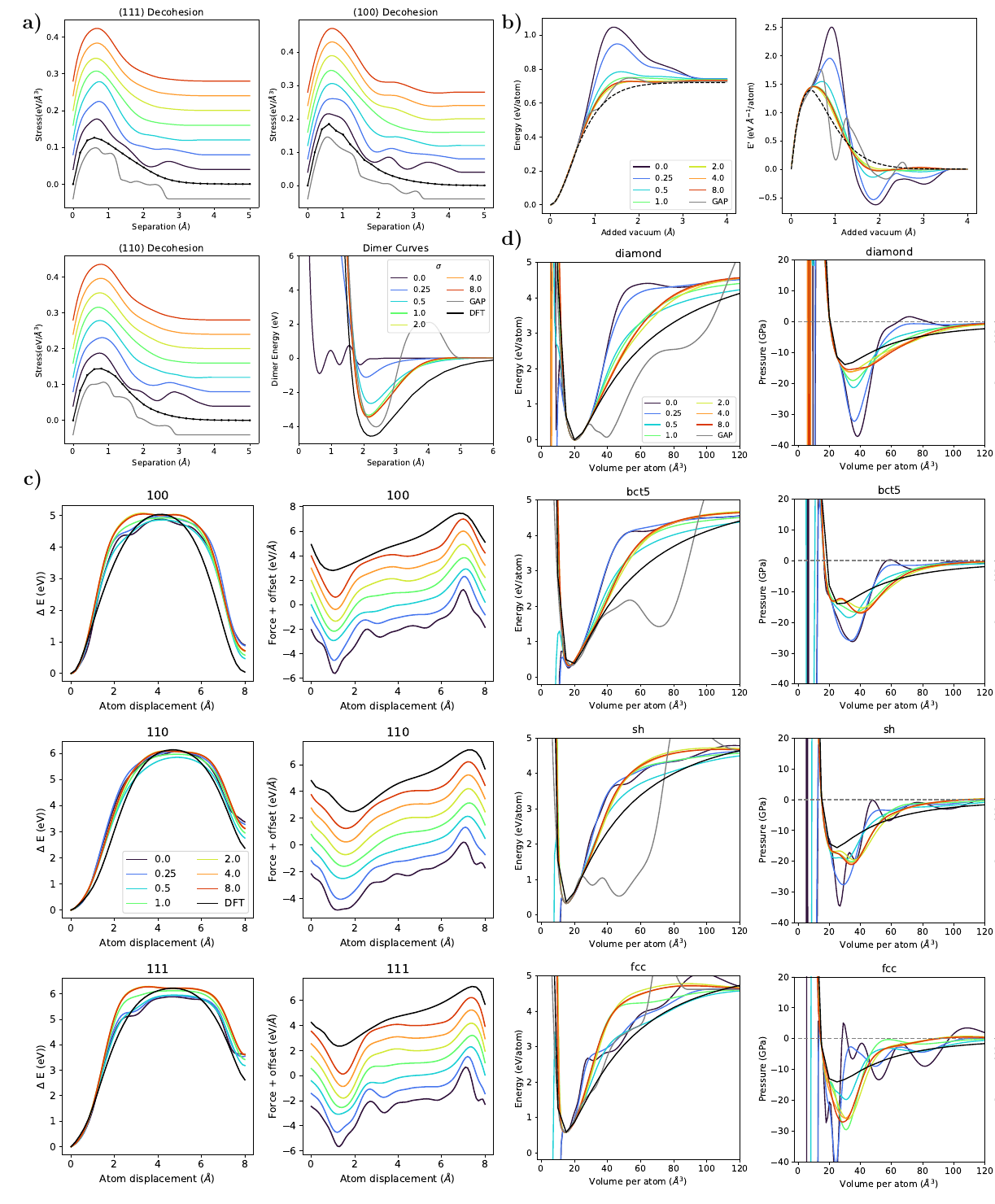}
    \caption{A summary of the tests described in the main text is shown for potentials fit to \textbf{Si10pc} using a \textbf{Gaussian} regularity prior. Specifically the diagrams show: a) the stress during the decohesion of Silicon diamond to various surfaces as well as the corresponding dimer curves. b) the effect of exfoliating of silicon diamond into silicene layers c) what happen when an atom is move across the vacuum from the lower surface of a silicon diamond slab to the upper surface d) the energy (left), and pressure (right) vs volume as various silicon polymorphs are expanded isotropically.}
    \label{fig:10_90_exponential}
\end{figure*}

\clearpage
\subsection{Full Silicon-GAP-18 dataset}
\begin{figure}[h!]
    \centering
    \includegraphics[width=0.5\textwidth]{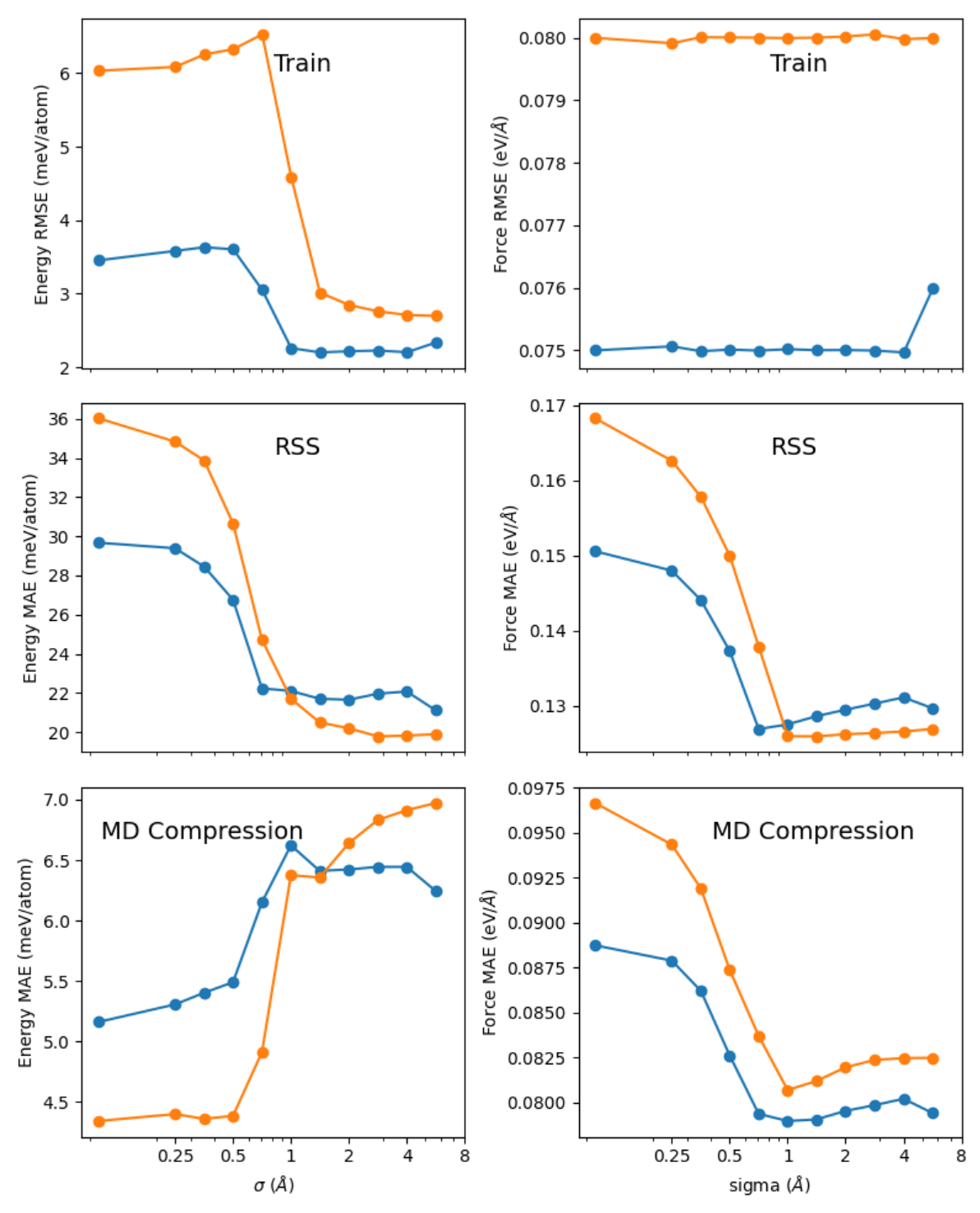}
    \caption{Energy and force errors on three different configuration types are shown for potentials fit to the full Silicon-GAP-18 dataset using a Gaussian regularity prior. The strength of the regularisation was chosen by tuning the force RMSE on the training set to be 0.06 eV/\AA. No test set error is shown as the entire dataset was used for training.}
    \label{fig:full_silicon_errors}
\end{figure}

\begin{figure*}
    \centering
    \includegraphics[width=\textwidth]{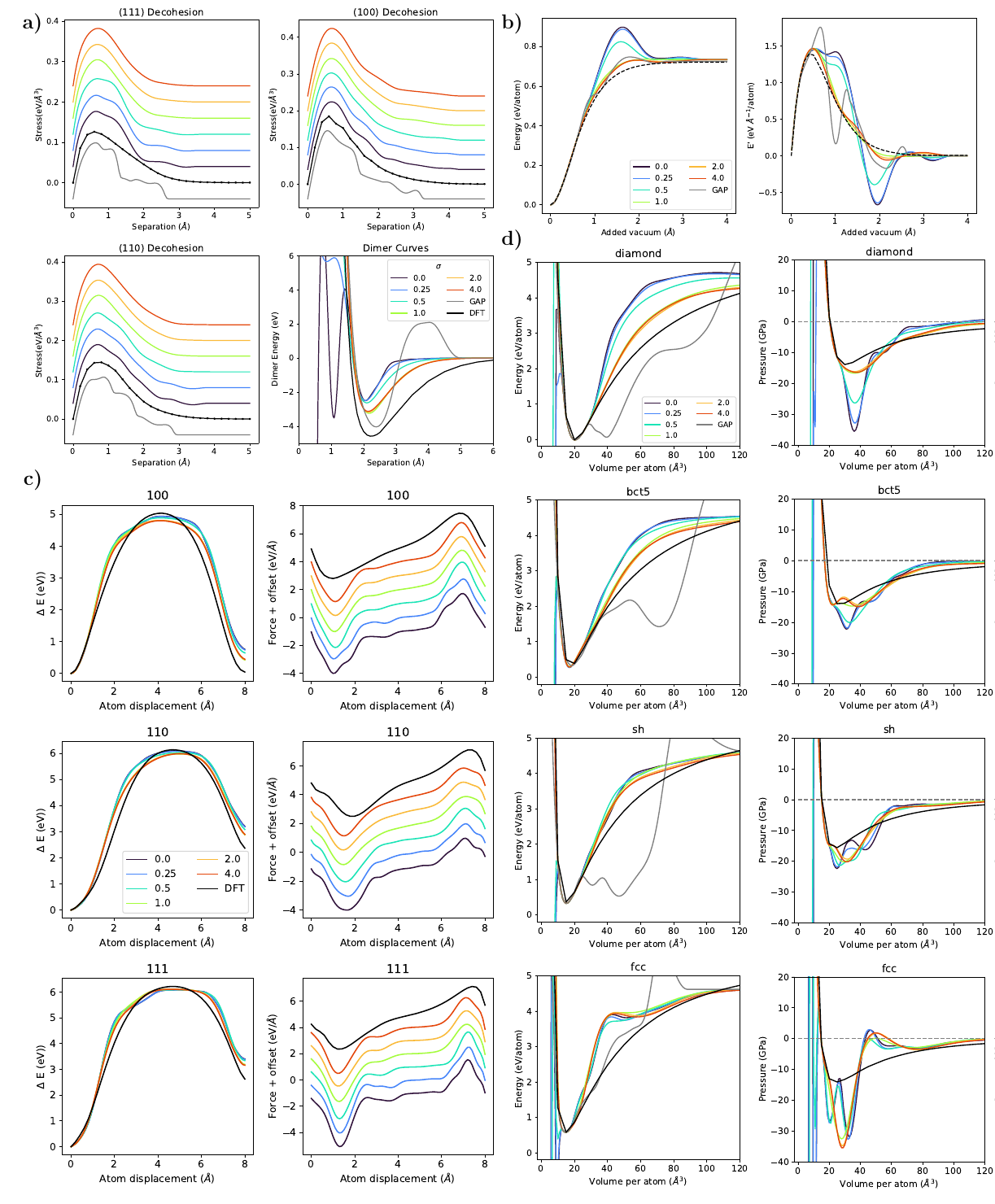}
    \caption{A summary of the tests described in the main text is shown for potentials fit to the \textbf{full Silicon-GAP-18 dataset} using a \textbf{Gaussian} regularity prior. a) stress during the decohesion of Silicon diamond to various surfaces and dimer curves. b) exfoliation of silicon diamond to silicene layers c) an atom is move across the vacuum from the lower surface of a silicon diamond slab to the upper surface d) energy (left), and pressure (right), vs volume as various silicon polymorphs are expanded isotropically.}
    \label{fig:full_dataset_gaussian}
\end{figure*}

\clearpage
\subsection{Aspirin}

\begin{figure}[h!]
    \centering
    \includegraphics[width=0.5 \textwidth]{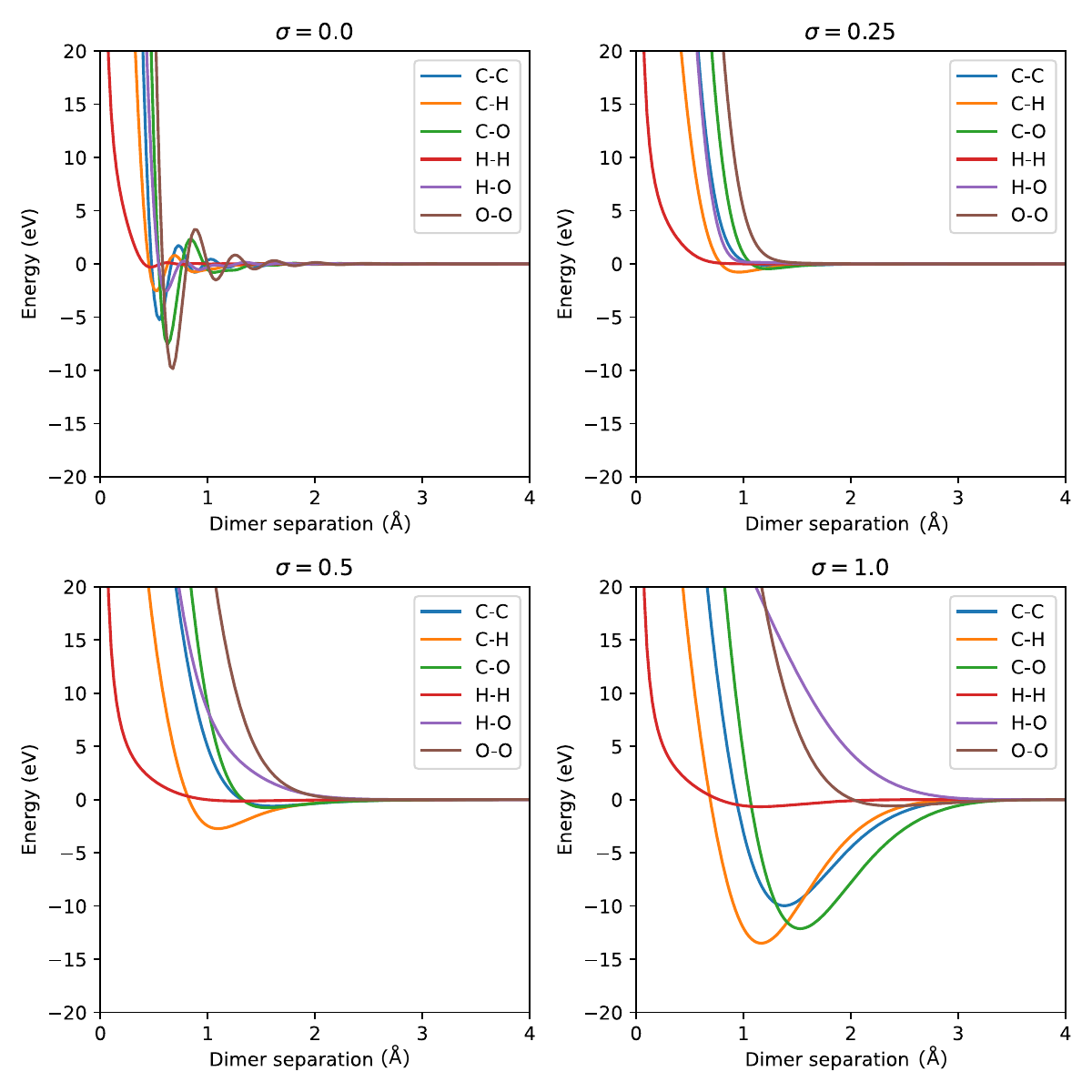}
    \caption{Dimer curves are shown for for ACE potentials fit to the Aspirin dataset using a Gaussian regularity prior with various values of $\sigma$. Only full Aspirin molecules (and the isolated atoms) are included in the training set so we do not expect these to accurately reflect the true \ac{DFT} PES. They are shown to indicate the effect of the regularity prior on the ACE PES.}
    \label{fig:aspirin_dimers}
\end{figure}

\begin{figure}
    \centering
    \includegraphics[width=0.5 \textwidth]{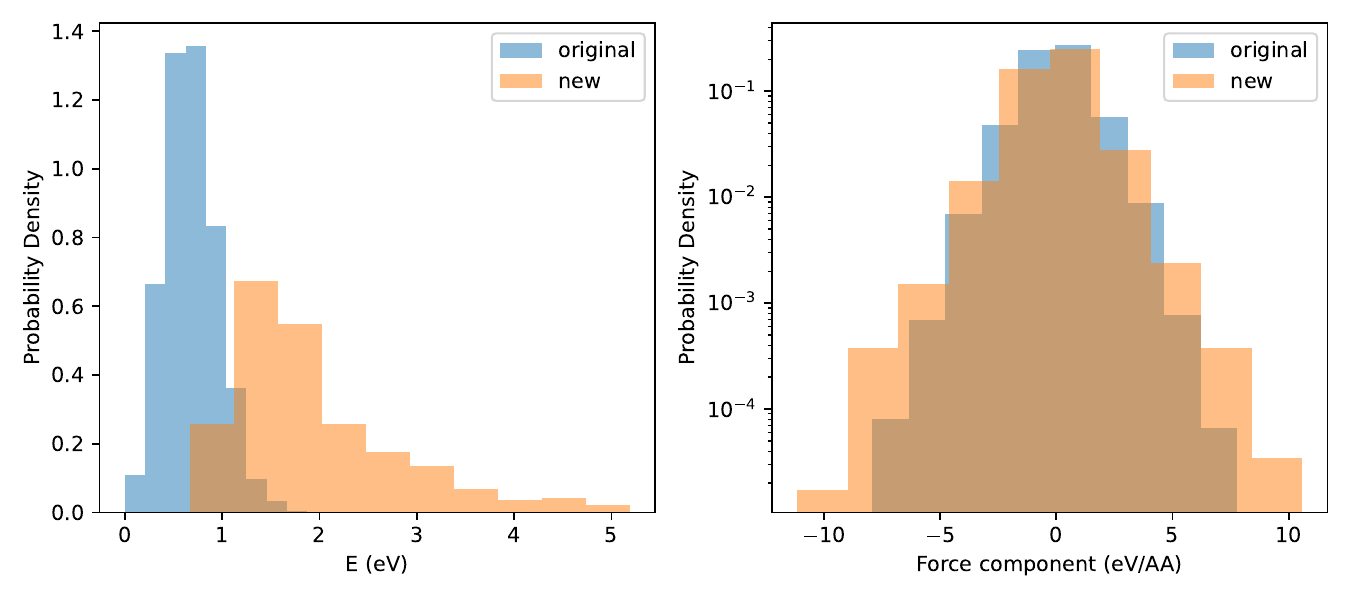}
    \caption{The distribution of energies and forces in the new Aspirin configurations generated via random dihedral rotations is compared to those in the original training and test sets.}
    \label{fig:aspirin_new_distribution}
\end{figure}

\begin{figure}
    \centering
    \includegraphics[width=0.5 \textwidth]{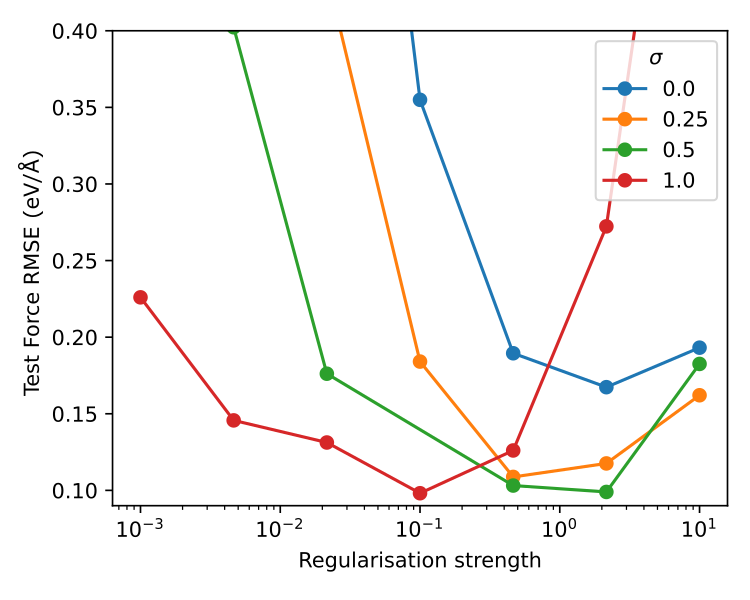}
    \caption{Force RMSE on the test set for ACE potentials fit to the new Aspirin training data. As the new test set also contains configurations generated via random dihedral rotations, the errors are expected to be larger;  see the distribution of forces in figure \ref{fig:aspirin_new_distribution}.}
    \label{fig:aspirin_EE_errors}
\end{figure}

\begin{figure}
    \centering
    \includegraphics[width=0.5\textwidth]{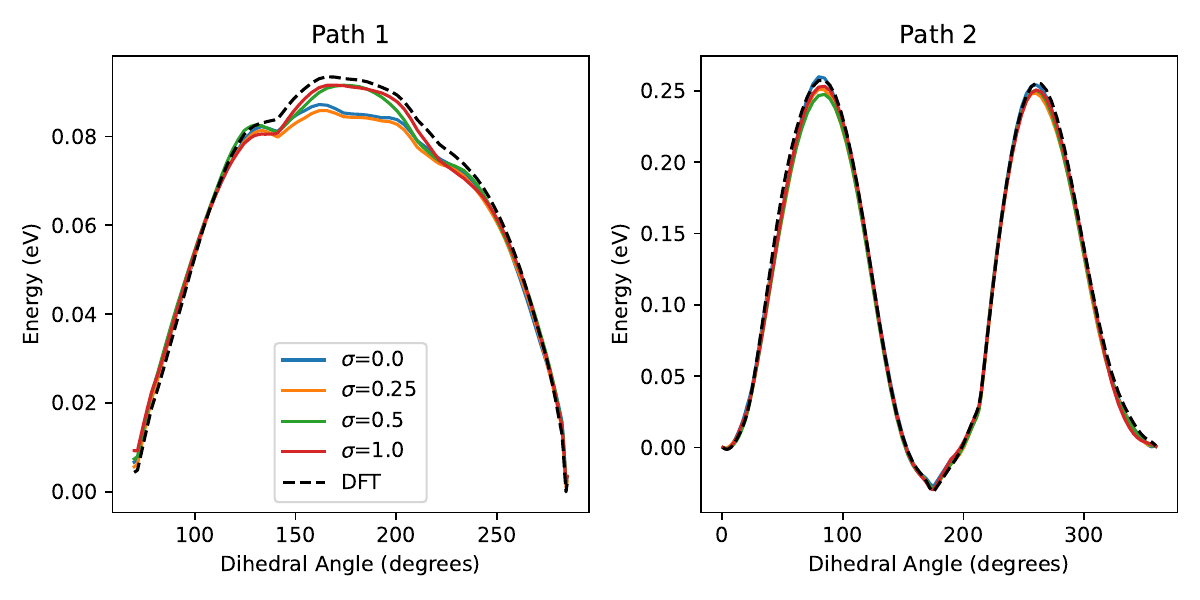}
    \caption{The relative energies along the NEB paths marked in figure\ref{fig:aspirin_MD_trajectories} are shown. All potential show good agreement with the \ac{DFT} reference indicating that the improved MD stability seen with larger values of $\sigma$ was not caused by fewer transitions between minima.}
    \label{fig:aspirin_neb_paths}
\end{figure}

\begin{figure}
    \centering
    \includegraphics[width=0.7\textwidth]{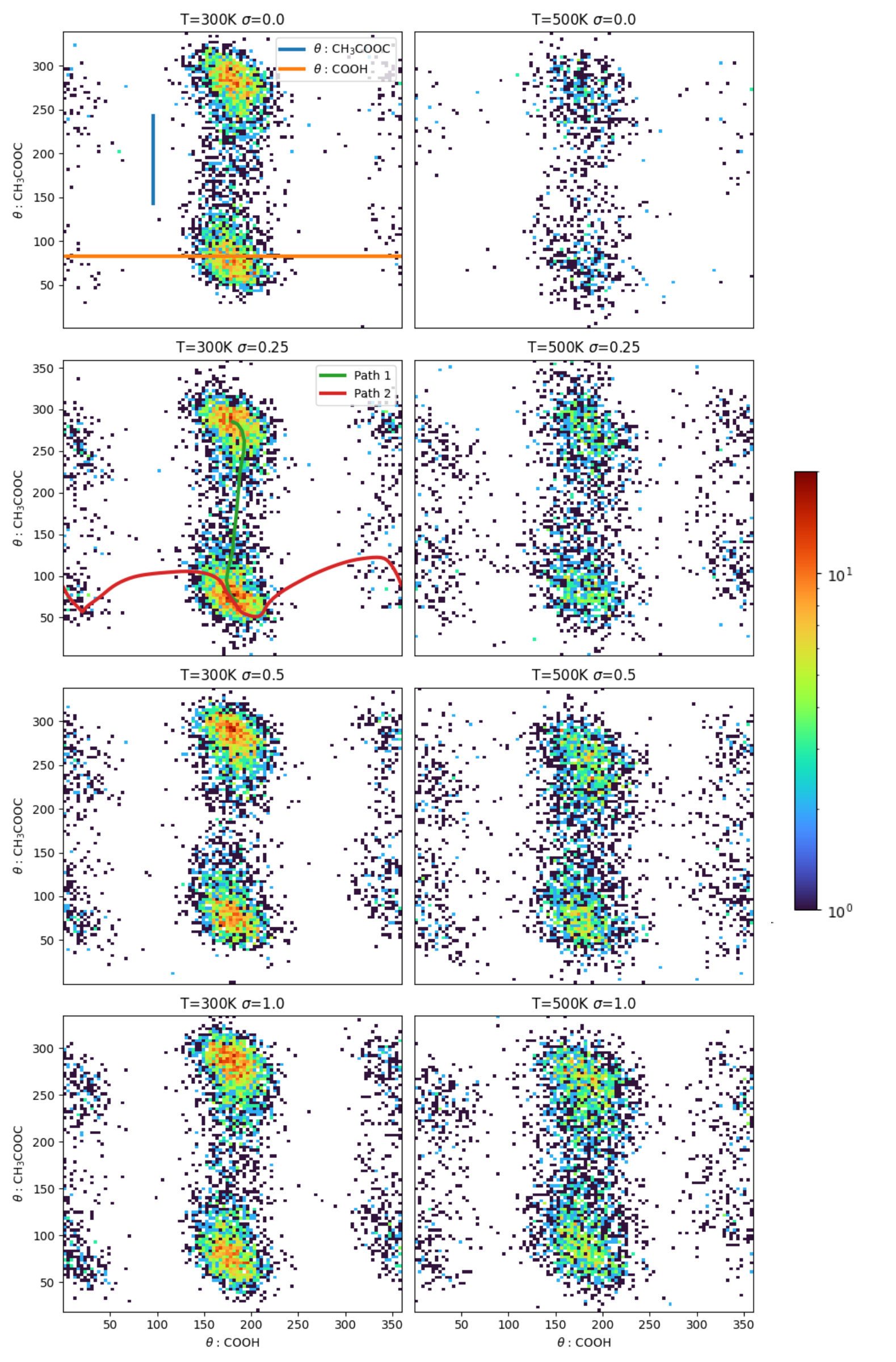}
    \caption{Distribution of dihedral angles from \ac{MD} performed using different ACE potentials. Where possible 100 equispaced frames were taken from every trajectory at a given T and $\sigma$; 100 trajectories per T, $\sigma$. Where fewer than 100 frames were collected all frames were used. The color map shows the log of the number density. The dihedral cuts shown in Figures \ref{fig:aspirin_dihedral_cuts} and \ref{fig:aspirin_EE_dihedrals}  are marked in the top left panel whilst two minimum energy paths (obtained using the Nudged Elastic Band method) between minima are marked in the panel below --- see figure\ref{fig:aspirin_neb_paths} for the corresponding energy profiles.}
    \label{fig:aspirin_MD_trajectories}
\end{figure}

\begin{figure}
    \centering
    \includegraphics[width=0.5 \textwidth]{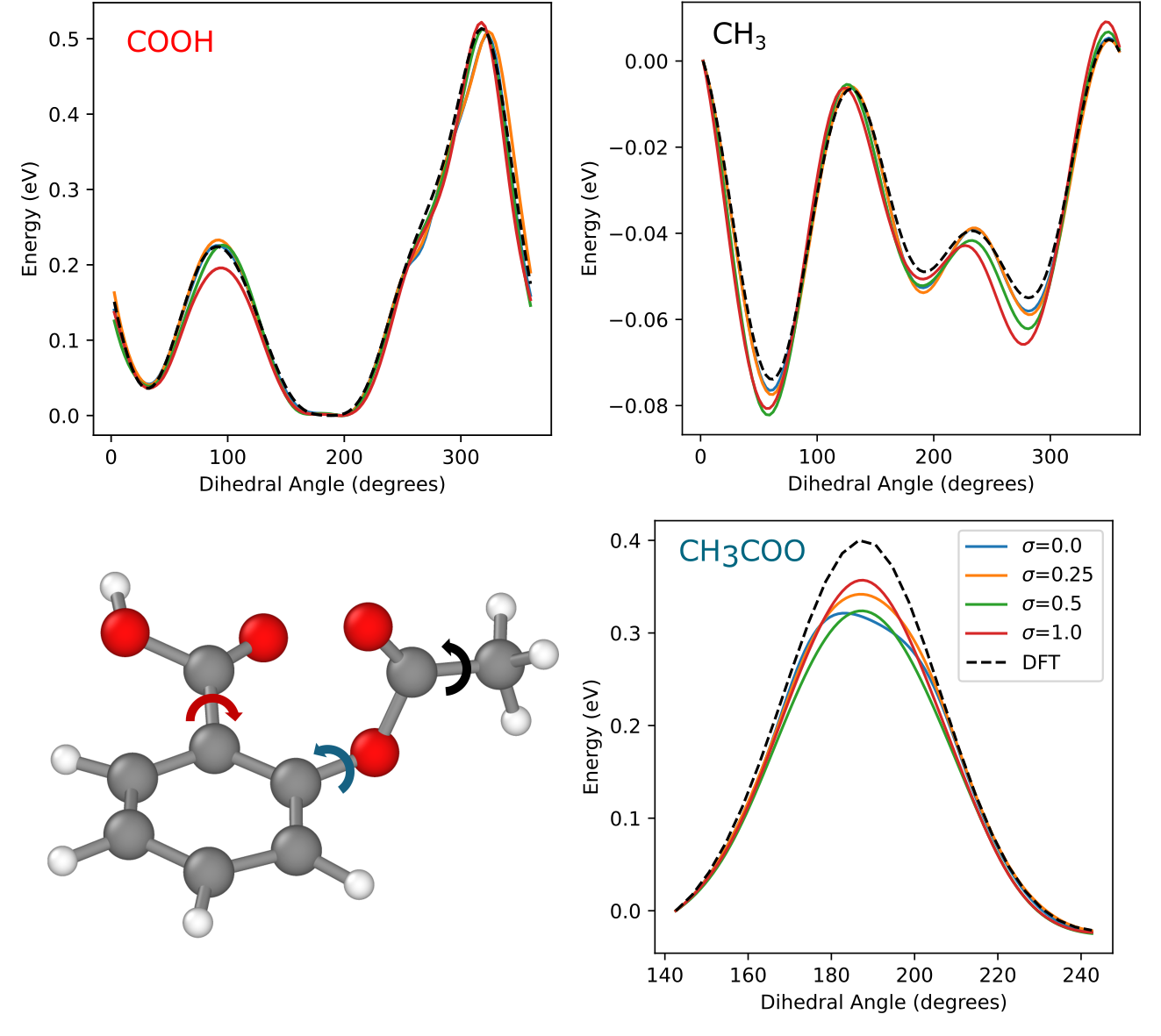}
    \caption{A single aspirin molecule is shown in the top right with three bonds marked. Anti-clockwise from the top left the energy of rigidly rotating the COOH (red), \ch{CH3COO} (blue) and \ch{CH3} (black) groups are shown as a function of angle for the ACE potentials fit to the enlarged training set containing configurations generated by applying random rotations about certain dihedral angles.}
    \label{fig:aspirin_EE_dihedrals}
\end{figure}

\end{document}